\documentclass[11pt]{article}
\usepackage{jcappub}
\usepackage{bm}
\usepackage{color}
\usepackage{array}
\usepackage{hhline}
\usepackage{graphicx}
\usepackage{multirow}
\usepackage{tikz,graphics,color,fullpage,float,epsf,caption,subcaption}
\usepackage{verbatim}
\usepackage{makecell}
\newcolumntype{P}[1]{>{\centering\arraybackslash}p{#1}}
\newcolumntype{M}[1]{>{\centering\arraybackslash}m{#1}}

\newcommand{\be}{\begin{equation}}
\newcommand{\ee}{\end{equation}}

\newcommand{\een}{\end{subequations}}
\newcommand{\ben}{\begin{subequations}}

\newcommand{\lsim}{\mathrel{\mathop{\kern 0pt \rlap
      {\raise.2ex\hbox{$<$}}}\lower.9ex\hbox{\kern-.190em $ \sim$}}}
\newcommand{\gsim}{\mathrel{\mathop{\kern 0pt
      \rlap{\raise.2ex\hbox{$>$}}}\lower.9ex\hbox{\kern-.190em $\sim$}}}

\newcommand{\CO}{\mathcal{O}}

\newcommand{\orcid}[1]{\href{https://orcid.org/#1}{\includegraphics[width=12pt]{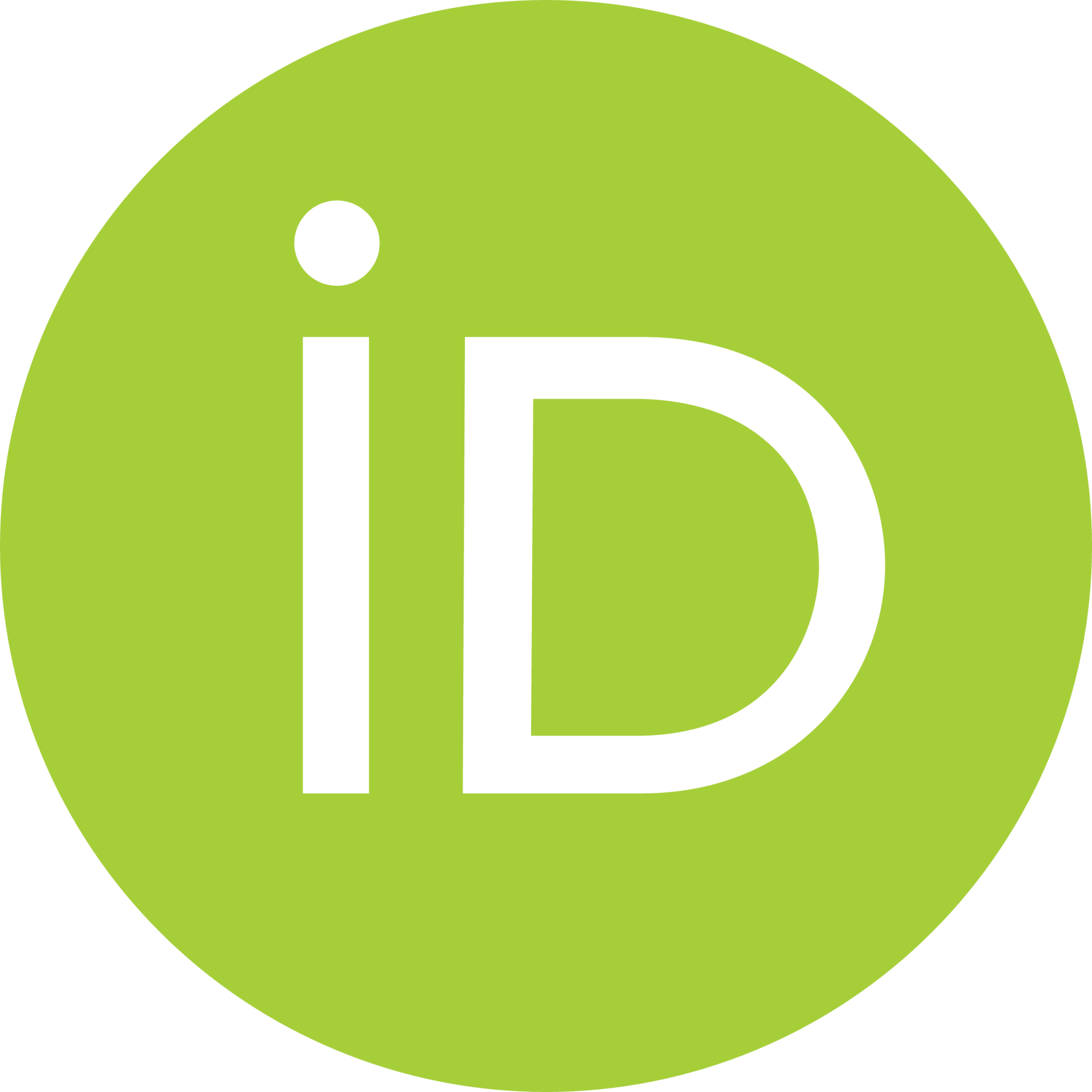}}}

\title{Low-mass constraints on WIMP effective models of inelastic scattering using the Migdal effect}

\author[a,c]{Sunghyun Kang\orcid{0000-0002-5241-1480},}
\author[a,c]{Stefano Scopel\orcid{0000-0002-9889-3091},}
\author[b]{Gaurav Tomar\orcid{0000-0002-3468-5306}}
\emailAdd{francis735@naver.com}
\emailAdd{scopel@sogang.ac.kr}
\emailAdd{tomar@iitp.ac.in}
\affiliation[a]{Department of Physics, Sogang University, 11 35 Baekbeom-ro, Mapo-gu, Seoul, 121-742, South Korea}
\affiliation[b]{Department of Physics, Indian Institute of Technology Patna, Bihar-801106, India}
\affiliation[c]{Center for Quantum Spacetime, Sogang University,
35 Baekbeom-ro, Mapo-gu, Seoul, 121-742, South Korea}
\abstract{
We use the Migdal effect to extend to low masses the bounds on each of the effective couplings of the non-relativistic effective field theory of a WIMP of mass $m_\chi$ and spin 1/2 that interacts inelastically with nuclei by either upscattering to a heavier state with mass splitting $\delta>0$ or by downscattering to a lighter state with $\delta<0$. In order to do so we perform a systematic analysis of the Migdal bounds in the $m_\chi-\delta$ parameter space comparing them to those from nuclear recoil searches.  
The Migdal effect allows to significantly extend to low WIMP masses the nuclear recoil bounds for $\delta<0$. In this case the bounds are driven by XENON1T, except when $\delta$ is vanishing or very small, when, depending on the WIMP--nucleus interaction, in the lower end of the $m_\chi$ range either DS50 or SuperCDMS are more constraining. On the other hand, when $\delta>0$ and the WIMP particle upscatters to a heavier state nuclear recoil bounds are stronger than those from the Migdal effect. 
}

\begin{document}
\hspace*{107.5mm}{CQUeST--2024--0738}\\
\maketitle

\section{Introduction}
\label{sec:introduction}

The absence of a signal from Weakly Interacting Massive Particles (WIMPs), the most popular dark matter (DM) candidates,  has prompted the scientific community to optimize direct detection experiments to the search of sub-GeV DM~\cite{LUX_2018,SENSEI:2020dpa, EDELWEISS:2020fxc, PandaX:2022ood, DAMIC:2019dcn, Essig:2011nj, Dolan_2017, migdal_vergados_1_2005, migdal_vergados_2_2005, Bell_2019, Cortona_2020, Wang_2021, Bell_2021_migdal, CDEX:2022kcd, Ibe:2017yqa, Berghaus:2022pbu, Adams:2022zvg}. One of the main challenges in detecting low-mass DM is the small energy deposited in the detector, which often lies below the detection thresholds. This makes it difficult to search for such DM particles, particularly through nuclear recoils. To overcome this challenge new analysis techniques such as the Migdal effect~\cite{migdal_1941} have been put forward. 
The Migdal effect allows to detect low-mass DM through secondary atomic ionization, and occurs when the WIMP--nucleus scattering process triggers the emission of an electron from the target~\cite{Ibe:2017yqa}.  
It has been first pointed out in~\cite{migdal_vergados_1_2005, migdal_vergados_2_2005, migdal_vergados_3_2005, dama_migdal} that if the DM particle couples comparably to both protons and electrons, the Migdal ionization rate can dominate over electron scattering for certain DM masses and mediator types, particularly in the hundreds of MeV range~\cite{Essig_2019,Baxter_2019}. Even lower masses can be probed in Migdal--type scenarios where the DM particle is a scalar that is absorbed by a nucleon or an electron triggering atomic ionization~\cite{Tan:2021nif, Dzuba:2023zcq}.

In direct detection experiments it is often easier to detect electromagnetic (EM) energy directly deposited in the measuring device compared to nuclear recoils, although it is more difficult to distinguish it from that generated from the background contamination. However, for very light WIMP masses the nuclear recoil process is kinematically not accessible, and the Migdal effect is the only way to detect a signal. 

Recent studies~\cite{Bell:2021zkr,Li:2022acp} have combined the Migdal effect with the Inelastic Dark Matter (IDM) scenario~\cite{inelastic}, where a WIMP state $\chi$ with mass $m_\chi$ interacts with nuclear targets by scattering to a different state $\chi^{\prime}$ with mass $m_\chi+\delta$. The  case $\delta<0$, indicated in the literature as exothermic DM~\cite{exothermic_dm}, implies the de-excitation of a meta-stable DM state that releases  additional energy in the detector, allowing to extend further the sensitivity to very light WIMPs.

In Refs.~\cite{Bell:2021zkr} and~\cite{Li:2022acp} the Migdal effect for IDM was analyzed considering  the cases of a spin-independent (SI) and spin-dependent (SD) WIMP--nucleus interaction, respectively. In our paper we wish to extend those analyses to a wider class of processes, by making use of the non-relativistic effective field theory (NREFT) framework~\cite{haxton1, haxton2}, that describes the most general WIMP--nucleus interactions for a particle of spin 1/2 by introducing a total of 14 momentum and velocity dependent operators (see Table~\ref{tab:operators}). 
Indeed, models that imply non--standard interactions for inelastic Dark Matter exist in the literature. For instance, in its non--relativistic limit Magnetic inelastic dark matter~\cite{Chang:2010en,Weiner:2012gm,Izaguirre:2015zva,Chatterjee:2022gbo} leads to the emergence of a combination of the ${\cal O}_1$, ${\cal O}_4$,  ${\cal O}_5$ and ${\cal O}_6$ operators, and in some kinematic conditions the ${\cal O}_5$ operator has been shown to dominate over ${\cal O}_1$ and to drive the scattering process (albeit this has been checked quantitatively only for the kinematics of elastic scattering~\cite{bishara_2017}).  So, on general grounds, one cannot rule out the possibility that, if inelastic DM exists, it interacts with nuclei in ways that are not standard.

Since previous studies discuss only specific benchmark values of the IDM parameters we conduct a thorough scan in terms of $\delta$ and $m_\chi$, showing that in absence of a signal the Migdal effect allows to exclude previously unexplored regions of the NREFT parameter space. 
In our analysis we employ the data from the XENON1T~\cite{XENON_migdal}, DS50~\cite{ds50_migdal}, and SuperCDMS~\cite{supercdms_migdal} experiments, revealing some degree of complementarity among them. 
The interplay between the Migdal effect and elastic scattering has been previously explored within the NREFT framework in~\cite{Bell_2019, Tomar:2022ofh}. 

Our paper is organized as follows: in Section~\ref{sec:migdal_inelastic} we summarize the Migdal effect within the context of IDM; our quantitative analysis is presented in Section~\ref{sec:analysis}; in Section~\ref{sec:conclusion} we provide our conclusions.

\section{Inelastic Scattering of Dark Matter under the Migdal Effect}
\label{sec:migdal_inelastic}
In this work we consider the inelastic scattering process $\chi T \rightarrow \chi^\prime T$ of a DM particle $\chi$ off a nuclear target $T$. Due to energy conservation,
\begin{equation}
    \frac{1}{2}\mu_{\chi T} v_T^2= E_{\chi^\prime}+E_T+\Delta,
    \label{eq:inelastic}
\end{equation}
where $\mu_{\chi T}=m_\chi m_T/(m_\chi+m_T)$ is the DM--nucleus reduced mass, $v_T\equiv|\vec{v}_T|$ is the DM particle incoming speed relative to the nucleus, that is assumed to be at rest in the lab rest frame. In the expression above $E_{\chi^\prime}$ and $E_T$ represent the energy of the outgoing DM particle and nucleus, respectively. Moreover, $\Delta$ is the amount of the initial kinetic energy of the DM particle lost due to the two simultaneous inelastic effects of the $\chi\rightarrow\chi^{\prime}$ transition and the target ionization with the emission of one electron. In particular, it is expressed as $\Delta=E_{EM}+\delta$, where $E_{EM}$ accounts for the electromagnetic energy deposited in the detector by the ionization process, and $\delta$ represents the mass splitting between the two DM states i.e. $\delta=m_{\chi^\prime}-m_\chi$ ($\delta <0$ corresponds to exothermic scattering and $\delta>0$ to endothermic scattering, respectively). 
It is clear from Eq.~(\ref{eq:inelastic}) that the maximum value of $\Delta$ is equal to the initial available kinetic energy of the incoming WIMP,
\begin{equation}
    \Delta_{\rm max}=\frac{1}{2}\mu_{\chi T} v_T^2.
\end{equation}
The nuclear recoil energy in the frame of the detector is expressed as,
\begin{equation}
    E_R=\frac{\mu_{\chi T}^2}{m_T}v^2\left[1-\frac{\Delta}{\mu_{\chi T}v^2}-{\rm cos}\theta\sqrt{1-\frac{2\Delta}{\mu_{\chi T}v^2}}~\right],
    \label{eq:atmoic_recoil}
\end{equation}
where $\theta$ is the DM--nucleon scattering angle in the center-of-mass frame. In Eq.~(\ref{eq:atmoic_recoil}), we made the approximation $\mu_{\chi^\prime T} \simeq \mu_{\chi T} $.

In our analysis we focus on the electron ionization rate in Xe, Ar, and Ge targets, and  make use of the ionization probabilities $p^c_{q_e}$ from~\cite{Ibe:2017yqa}, obtained under the single-atom approximation~\footnote{Recently, the calculation of the ionization probabilities of~~\cite{Ibe:2017yqa} was improved using the Dirac-Hartree-Fock method~\cite{Cox:2022ekg}. In the ionization energy ranges relevant to our analysis both calculations lead to similar results.}. In particular, the ionisation event rate in a direct detection experiment due to the Migdal effect is given by~\cite{Ibe:2017yqa},
\begin{equation}
\label{eq:diff_rate_migdal0}
 \frac{dR}{dE_{det}}=\int_0^{\infty} d E_R \int_{v_{min}(E_R)}^\infty dv_T \frac{d^3 R_{\chi T}}{dE_R dv_T dE_{det}} ,
\end{equation}
\noindent with
\begin{equation}
 \label{eq:diff_rate_migdal}
 \frac{d^3R_{\chi T}}{dE_Rdv_TdE_{det}}=\frac{d^2R_{\chi T}}{dE_R dv_T}\times \frac{1}{2\pi}\sum_{n,l}\frac{d}{dE_e}p^c_{q_e}(nl\rightarrow(E_e)).
 \end{equation}
In our analysis we include the electrons up to the shells $n$ = 3 (SuperCDMS and DS50), 4, 5 (XENON1T).
It is worth noticing  that in the lab frame the ionization probabilities $p^c_{q_e}$ are boosted due to the recoil of the nucleus, and scale linearly with $E_R$~\cite{Ibe:2017yqa}. 
The Migdal effect is only relevant at low WIMP masses and becomes subdominant as soon as nuclear recoils are kinematically allowed. In the expression above $E_{det}$ represents the total deposited energy and is given by,

\begin{eqnarray}
E_{det}&=&QE_R+E_{EM}+\delta=QE_R+\Delta,\nonumber\\
    E_{EM}&=&E_e+E_{nl},
\end{eqnarray}

\noindent where $Q$ is the quenching factor, $E_e$ represents the energy of the outgoing electron and $E_{nl}$ the atomic de-excitation energy. In particular for $\Delta\le0$ the electromagnetic energy $QE_R$ produced by the recoil of the nucleus is always negligible compared to that produced by the Migdal process, leading to $E_{det}\simeq E_{EM}+\delta$. The minimum and the maximum of the recoil energy $E_R$ are obtained from Eq.~(\ref{eq:atmoic_recoil}), using $\theta=0$ and $\pi$ respectively. Finally, the minimum DM velocity required for the recoil of the nucleus at a given energy $E_R$ is given by,
\begin{equation}
    v_{min}(E_R)=\frac{1}{\mu_{\chi T}\sqrt{2m_T E_R}}|m_{T} E_R+\mu_{\chi T} \Delta|.
    \label{eq:vmin}
\end{equation}
In our analysis, we adopt a standard Maxwell--Boltzmann velocity distribution at rest in the Galactic frame with escape velocity $v_{esc}=550$ km/s and boosted to the solar rest frame with rotation speed of  $v_0=220$ km/s. 

As a result of the small momentum transfer the WIMP--nucleus scattering process is non relativistic and can be described in terms of the most general NREFT allowed by symmetry under Galilean boosts~\cite{haxton1,haxton2}. In this context, the expression for the interaction Hamiltonian between the target nucleus, and the WIMP particle is given by,
\begin{eqnarray}
{\bf\mathcal{H}}({\bf{r}})&=& \sum_{\tau=0,1} \sum_{j=1}^{15} c_j^{\tau} \mathcal{O}_{j}({\bf{r}}) \, t^{\tau} ,
\label{eq:H}
\end{eqnarray}

\noindent with $t^0=1$ and $t^1=\tau_3$ the $2\times2$ identity and third Pauli matrix in isospin space, respectively. The isoscalar and isovector (dimension -2)
coupling constants $c^0_j$ and $c^{1}_j$, are related to those to protons and neutrons $c^{p}_j$ and $c^{n}_j$ by $c^{p}_j=(c^{0}_j+c^{1}_j)/2$ and $c^{n}_j=(c^{0}_j-c^{1}_j)/2$.  For a spin-1/2 WIMP there are 14 operators $\mathcal{O}_j$~\cite{haxton1,haxton2,all_spins}, which depend at most linearly on the transverse velocity $v^\perp$ (defined as the component of $\vec{v}_T$ perpendicular to the transferred momentum $\vec{q}$), and that are listed in Table~\ref{tab:operators}. In the case of the standard SI and SD interactions (corresponding to the $\mathcal{O}_1$ and $\mathcal{O}_4$ operators, respectively, in the convention of~\cite{haxton1,haxton2}) it is common to express the couplings $c_1^N$ and $c_4^N$ (with $N=p,n$) in terms of the corresponding WIMP--nucleon  scattering cross-sections,
\begin{eqnarray}\nonumber
    \sigma_{\rm SI}^N &=& \frac{(c^N_1)^2\mu^2_{\chi N}}{\pi},\\ 
    \sigma_{\rm SD}^N &=& \frac{3}{16}\frac{(c^N_4)^2\mu^2_{\chi N}}{\pi},
    \label{eq:si_sd_cs}
\end{eqnarray}

\noindent where $\mu_{\chi N}$ is the WIMP--nucleon reduced mass.

The detailed expression for the calculation of the differential rate $\frac{d^2R_{\chi T}}{dE_R dv_T}$ in Eq.~(\ref{eq:diff_rate_migdal}) is provided in Section 2 of \cite{sogang_scaling_law_nr}, which has been implemented in the WimPyDD code~\cite{wimpydd}. Such expression scales linearly with the DM density in the neighborhood of the Sun, for which we adopt the standard value $\rho_\odot$ = 0.3 GeV/cm$^3$. Moreover, we parameterize with $f$ the fraction of $\rho_\odot$ provided by the particle $\chi$ (notice that DD experiments are only sensitive to the combination $\sqrt{f}\times c_{j}$).

In the non-relativistic limit the WIMP--nucleus differential cross
section is proportional to the squared amplitude,

\be
\frac{d\sigma_T}{d E_R}=\frac{2 m_T}{4\pi v_T^2}\left [ \frac{1}{2 j_{\chi}+1} \frac{1}{2 j_{T}+1}|\mathcal{M}_T|^2 \right ],
\label{eq:dsigma_de}
\ee

\noindent with $m_T$ the nuclear mass, $j_T$, $j_\chi$ the spins of the target nucleus and of the WIMP, where $j_\chi=1/2$, and \cite{haxton2},

\begin{equation}
  \frac{1}{2 j_{\chi}+1} \frac{1}{2 j_{T}+1}|\mathcal{M}_T|^2=
  \frac{4\pi}{2 j_{T}+1} \sum_{\tau=0,1}\sum_{\tau^{\prime}=0,1}\sum_{k} R_k^{\tau\tau^{\prime}}\left [c^{\tau}_i,c_j^{\tau^{\prime}},(v^{\perp})^2,\frac{q^2}{m_N^2}\right ] W_{T k}^{\tau\tau^{\prime}}(y).
\label{eq:squared_amplitude}
\end{equation}

\noindent In the expression above the squared amplitude
$|\mathcal{M}_T|^2$ is summed over initial and final spins and the
$R_k^{\tau\tau^{\prime}}$'s are WIMP response functions provided, for instance, in Eq. (38) of~\cite{haxton2}. They depend
on the couplings $c^{\tau}_j$ as well as the transferred momentum, while,

\begin{equation}
(v^{\perp})^2=v^2_T-v_{min}^2,
\label{eq:v_perp}
\end{equation}

\noindent and $v_{min}$ is given by Eq.~(\ref{eq:vmin}).
Moreover, in Eq.~(\ref{eq:squared_amplitude}) the $W^{\tau\tau^{\prime}}_{T k}(y)$'s are nuclear response functions and the index $k$ represents different effective nuclear operators, which, under the assumption that the nuclear ground state is an approximate eigenstate of $P$ and $CP$, can be at most eight: following the notation in \cite{haxton1,haxton2}, $k$=$M$, $\Phi^{\prime\prime}$, $\Phi^{\prime\prime}M$, $\tilde{\Phi}^{\prime}$, $\Sigma^{\prime\prime}$, $\Sigma^{\prime}$, $\Delta$, $\Delta\Sigma^{\prime}$. The $W^{\tau\tau^{\prime}}_{T k}(y)$'s are function of $y\equiv (qb/2)^2$, where $b$ is the size of the nucleus. For the target nuclei $T$ used in most direct detection experiments the functions $W^{\tau\tau^{\prime}}_{T k}(y)$, calculated using nuclear shell models, have been provided in Refs.~\cite{haxton2,catena}.

\begin{table}[t!]
\begin{center}
\begin{tabular}{|l|l|}
\hhline{|-|-|}
$ \CO_1 = 1_\chi 1_N$ & $\CO_9 = i \vec{S}_\chi \cdot (\vec{S}_N \times \frac{\vec{q}}{m_N})$ \\
$\CO_3 = i \vec{S}_N \cdot (\frac{\vec{q}}{m_N} \times \vec{v}^\perp)$ & $\CO_{10} = i \vec{S}_N \cdot \frac{\vec{q}}{m_N}$ \\
$\CO_4 = \vec{S}_\chi \cdot \vec{S}_N$ & $\CO_{11} = i \vec{S}_\chi \cdot \frac{\vec{q}}{m_N}$\\
$\CO_5 = i \vec{S}_\chi \cdot (\frac{\vec{q}}{m_N} \times \vec{v}^\perp)$ & $\CO_{12} = \vec{S}_\chi \cdot (\vec{S}_N \times \vec{v}^\perp)$ \\
$\CO_6= (\vec{S}_\chi \cdot \frac{\vec{q}}{m_N}) (\vec{S}_N \cdot \frac{\vec{q}}{m_N})$ & $\CO_{13} =i (\vec{S}_\chi \cdot \vec{v}^\perp  ) (  \vec{S}_N \cdot \frac{\vec{q}}{m_N})$ \\
$\CO_7 = \vec{S}_N \cdot \vec{v}^\perp$ & $\CO_{14} = i ( \vec{S}_\chi \cdot \frac{\vec{q}}{m_N})(  \vec{S}_N \cdot \vec{v}^\perp )$\\
$\CO_8 = \vec{S}_\chi \cdot \vec{v}^\perp$ & $\CO_{15} = - ( \vec{S}_\chi \cdot \frac{\vec{q}}{m_N}) \big((\vec{S}_N \times \vec{v}^\perp) \cdot \frac{\vec{q}}{m_N}\big)$ 
\\ \hline
\end{tabular}
\caption{Non-relativistic Galilean invariant operators for dark matter with spin $1/2$.}
\label{tab:operators}
\end{center}
\end{table}

The ionization probabilities $p_{q_e}^c$ are calculated using the impulse approximation, that assumes that the DM--nucleus collisions happen rapidly compared to the time scale set by the potential well of the detector lattice, $1/\omega_{ph}$, with $\omega_{ph}$ the phonon frequency~\cite{Knapen_2020}. In particular, the time scale of DM--nucleus collision and the emission of Migdal electrons is of the order $t\simeq 1/E_R$ (in natural units). As a consequence, the validity of the impulse approximation implies a lower cut on $E_R$. 

In the case of a xenon fluid, we utilise a time scale of $t=10^{-12}$ s, which applies to xenon at 170K~\cite{Bell:2021zkr}. Consequently, in the energy integration of Eq.~(\ref{eq:diff_rate_migdal0}) we set $E_R > E_{cut}=100 t^{-1} \approx 50$ meV. Using Eq.~(\ref{eq:atmoic_recoil}) for elastic scattering this implies $m_\chi \geq 0.02$ GeV. However, through the $\Delta$ parameter the lower bound on $m_\chi$ depends on the mass splitting $\delta$ and on the experimental threshold on $E_{det}$. For instance, taking into account the XENON1T energy threshold (see Section~\ref{sec:analysis}) for $\delta=-10$ keV, one gets $m_\chi \geq 0.6$ MeV. 

For an argon fluid at 90K using an inter-atomic distance of 4 Angstroms and a sound speed of 847.55 m/s~\cite{bowman1968velocity}, we estimate a time scale of $t=a/v_s \approx 10^{-12}$ s leading again to $E_{cut}\simeq$ 50 meV. In the case of elastic scattering, this implies $m_\chi\geq 0.01$ GeV. 
For $\delta=-10$ keV, we set $m_\chi \geq 0.2$ MeV.

As far as the germanium target is concerned, we closely follow the work of~\cite{Knapen_2020}, where it is shown that the impulse approximation holds if the momentum transfer $q_T$ satisfies the condition $q_T \geq \sqrt{2m_T \omega_{ph}}$, with $\omega_{ph}$ estimated between 30--50 meV for a typical crystal. Taking a conservative approach, we assume $E_{cut} = 50$ meV. Using this criterion, we find $m_\chi \geq 0.016$ GeV for elastic scattering, 
while for $\delta=-10$ keV, we obtain $m_\chi \geq 0.34$ MeV.

In the case of endothermic scattering ($\delta>0$) rather than from the impulse approximation cut the lower bound on $m_\chi$ is obtained from kinematics, specifically from the requirement that the argument of the square root in Eq.~(\ref{eq:atmoic_recoil}) is positive. For instance, for $\delta=10$ keV this leads to $m_\chi \gtrsim 3$ GeV  (independent on the target as long as $m_\chi\ll m_T$).
\begin{figure}
    \centering
        \centering
        \includegraphics[width=0.8\linewidth]{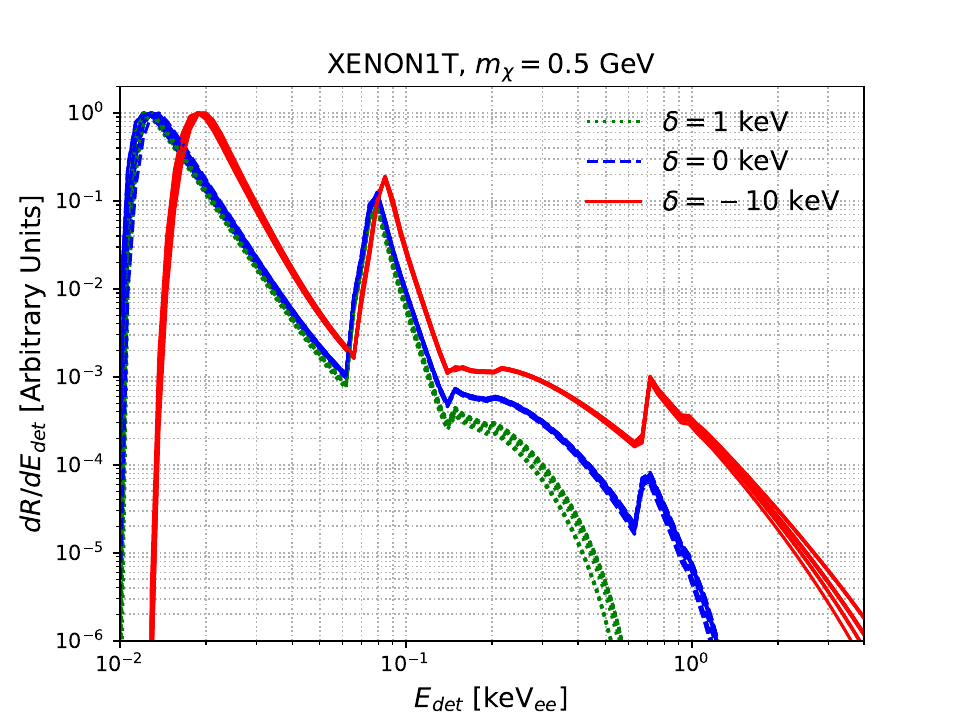}
    \caption{
The normalized Migdal spectrum is obtained from DM scattering with the xenon target for $m_\chi=0.5$ GeV, considering $\delta=1$ (dotted green), 0 (dashed blue), and $-10$ (solid red) keV, incorporating all the 14 interaction couplings of Eq.~(\ref{eq:H}). The NREFT interactions have a mild effect on the energy spectrum and are determined by the ionization probabilities (see text for details). Similar conclusions apply to argon and germanium targets.}
    \label{fig:diff_rate_migdal_comb}
\end{figure}
\section{Analysis}
\label{sec:analysis}
The Migdal event rate is obtained by multiplying the WIMP--nucleus scattering rate $\frac{d^2R_{\chi T}}{dE_R dv_T}$ with the ionization probability $p^c_{q_e}$ (Eq.~(\ref{eq:diff_rate_migdal})). As shown in Eq.~(\ref{eq:vmin}), the emission of an electron leads to modifications in the kinematics of the scattering rate $R$ compared to elastic scattering. In particular, Eq.~(\ref{eq:vmin}) mimics that for the kinematics of IDM~\cite{inelastic}, with the replacement of the mass splitting $\delta$ with $\Delta$, defined as $\Delta = E_{EM} + \delta$. 

In WimPyDD~\cite{wimpydd} we implement the kinematics of the Migdal effect through the argument \verb|delta| of the \verb|wimp_dd_rate| routine. Specifically, we adopted $\verb|delta|=E_{EM}+\delta \simeq E_e +\delta$, and integrated the scattering rate over the undetected nuclear scattering energy $E_R>E_{cut}\simeq$50 meV, in order to comply with the impulse approximation as discussed in the previous Section. Furthermore, we used the ionization probability $p^c_{q_e}$ for Xe, Ar, and Ge atoms from~\cite{Ibe:2017yqa}, corresponding to the XENON1T~\cite{XENON_migdal}, DS50~\cite{ds50_migdal}, and SuperCDMS~\cite{supercdms_migdal} experiments, respectively. 

In Fig.~\ref{fig:diff_rate_migdal_comb} we provide an explicit example of the energy dependence of the Migdal differential rate in the case of different NREFT interactions, for a Xe target and $m_\chi=0.5$ GeV. In all the plots the curves have 
the same normalization, to show that, especially in the energy range that drives the overall signal, the NREFT interaction has a mild effect on the energy spectrum, which is mainly determined by the nuclear ionization probabilities. The solid red, dashed blue, and dotted green lines represent $\delta=-10$ keV (exothermic), $\delta=0$ keV, and $\delta=1$ keV (endothermic), respectively, for all the 14 interaction couplings of Eq.~(\ref{eq:H}).  Different peaks in Fig.~\ref{fig:diff_rate_migdal_comb} correspond to the contributions of the shells $n=3,4,5$. It is worth mentioning that depending on the thresholds of the considered experiments, only some of the shells contribute to the Migdal event rate.

\begin{figure}
    \centering
    \begin{subfigure}{0.49\linewidth}
        \centering
        \includegraphics[width=1\linewidth]{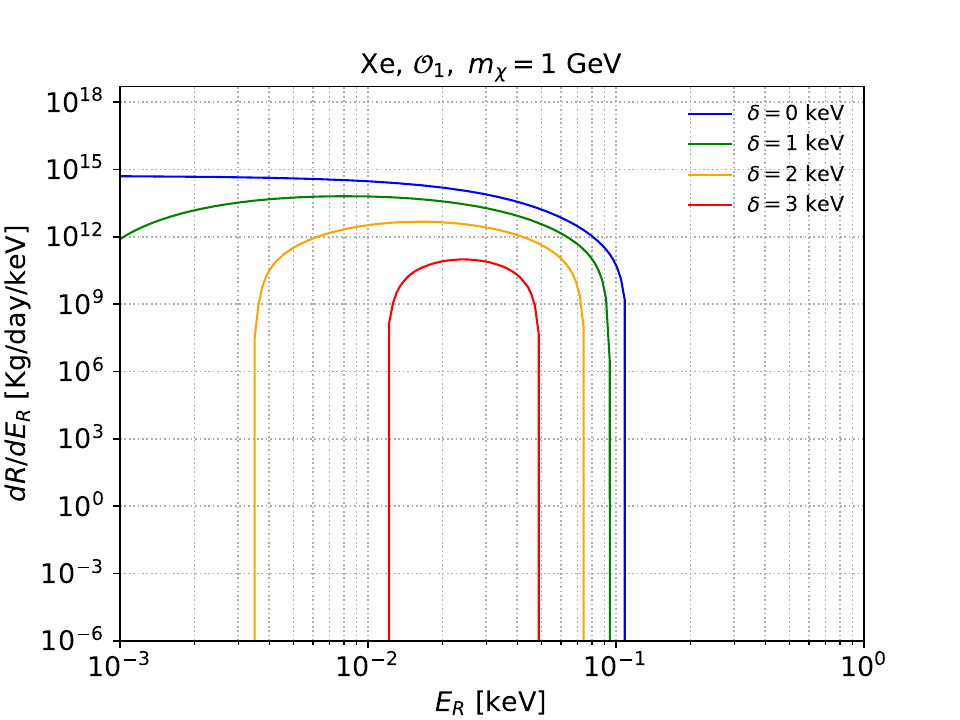}
    \end{subfigure}
    \begin{subfigure}{0.49\linewidth}
        \centering
        \includegraphics[width=1\linewidth]{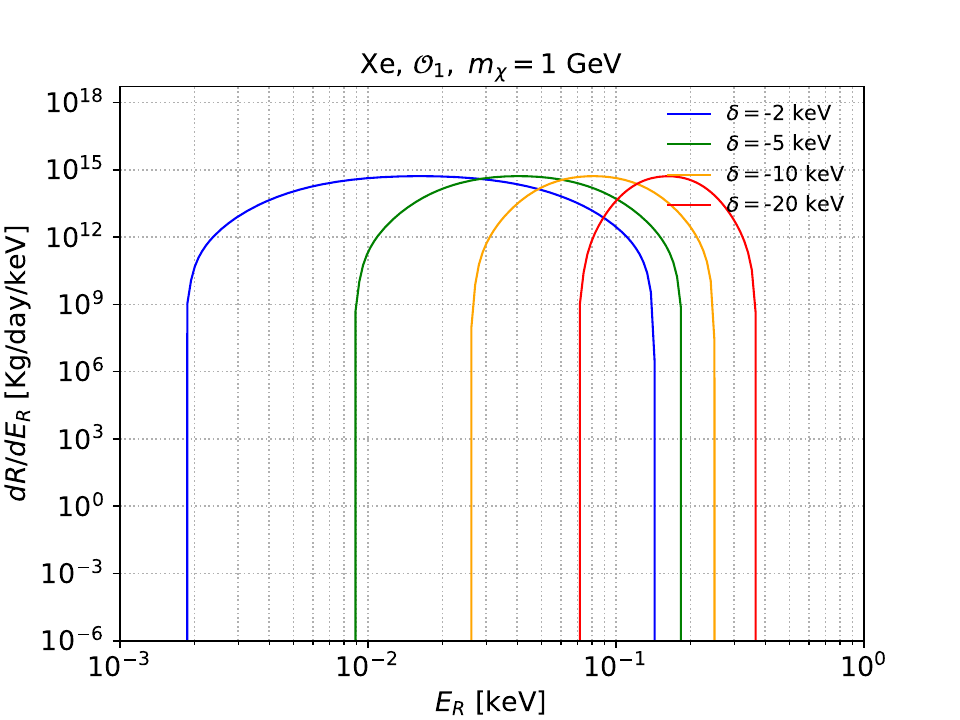}
    \end{subfigure}
    \caption{The nuclear recoil spectrum for $\delta<0$ (exothermic), $\delta=0$, and $\delta>0$ (endothermic) scattering of DM with a Xe target, assuming $\mathcal{O}_1$ interaction with a DM mass of $m_\chi=1$ GeV.}
    \label{fig:diff_rates_1}
\end{figure}

In Fig.~\ref{fig:diff_rates_1} we plot the nuclear recoil spectrum for a Xe target, taking $m_\chi = 1$ GeV. From such plot one can observe that for endothermic DM ($\delta>0$) the normalization of the spectrum decreases with $\delta$ while at the same time the energy interval where the spectrum is different from zero shrinks. This leads to a suppression of the Migdal signal. On the other hand for exothermic scattering ($\delta < 0$) both the minimum and the maximum nuclear recoil energies are shifted to higher values for growing $|\delta|$, while the normalization of the spectrum remains the same. This leads to an enhancement for the Migdal signal because, as previously pointed out, the ionization probabilities grow linearly with $E_R$ due to the boost from the recoiling atom to the lab frame. Similar conclusions hold for other NREFT interactions and targets, so that, on general grounds, compared to the elastic case the Migdal signal is enhanced for $\delta<0$ and suppressed for $\delta>0$. 
It is again important to mention that the shape of the Migdal energy spectrum is determined by the ionization probabilities $p^c_{q_e}$ and remains approximately the same irrespective of the interaction operator, of $m_\chi$ and of $\delta$.

\begin{figure*}
\begin{center}
    \includegraphics[width=0.44\textwidth]{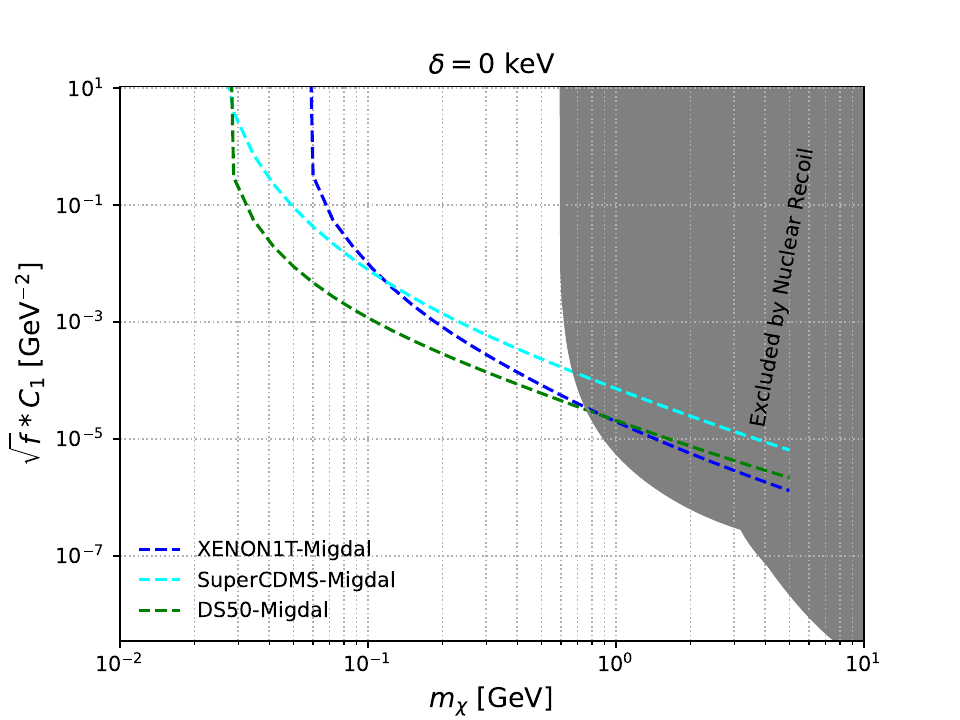}
    \includegraphics[width=0.44\textwidth]{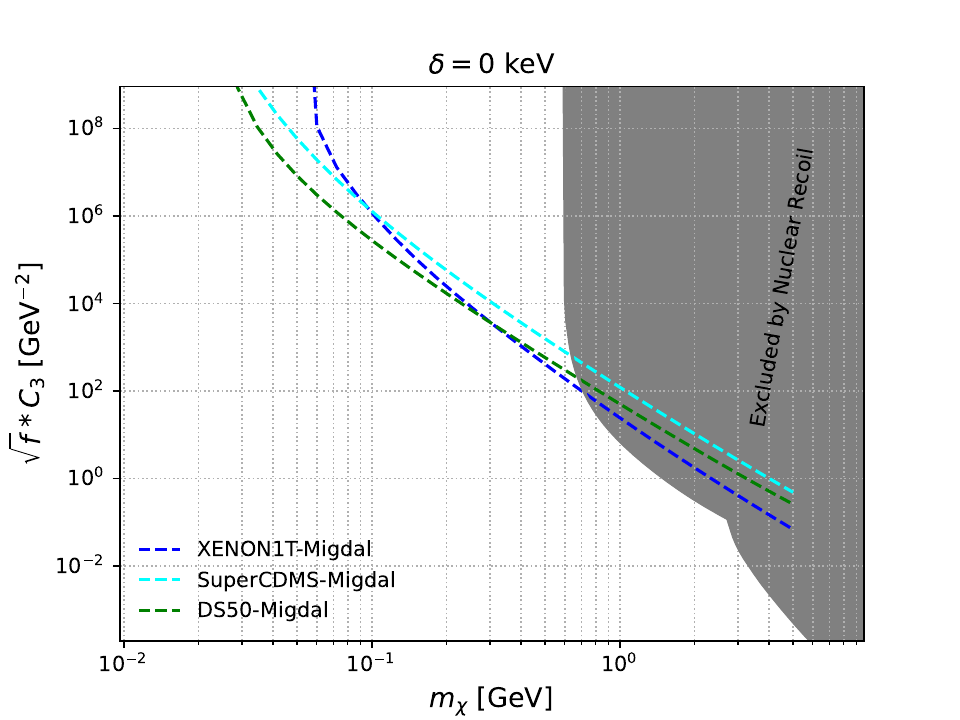}\\
    \includegraphics[width=0.44\textwidth]{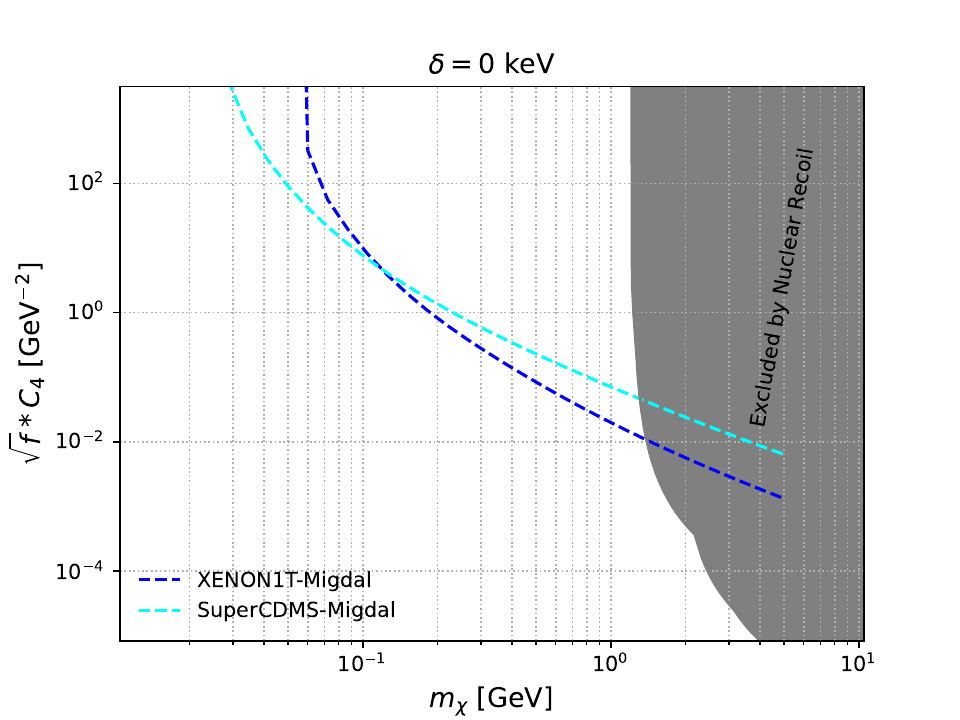}
    \includegraphics[width=0.44\textwidth]{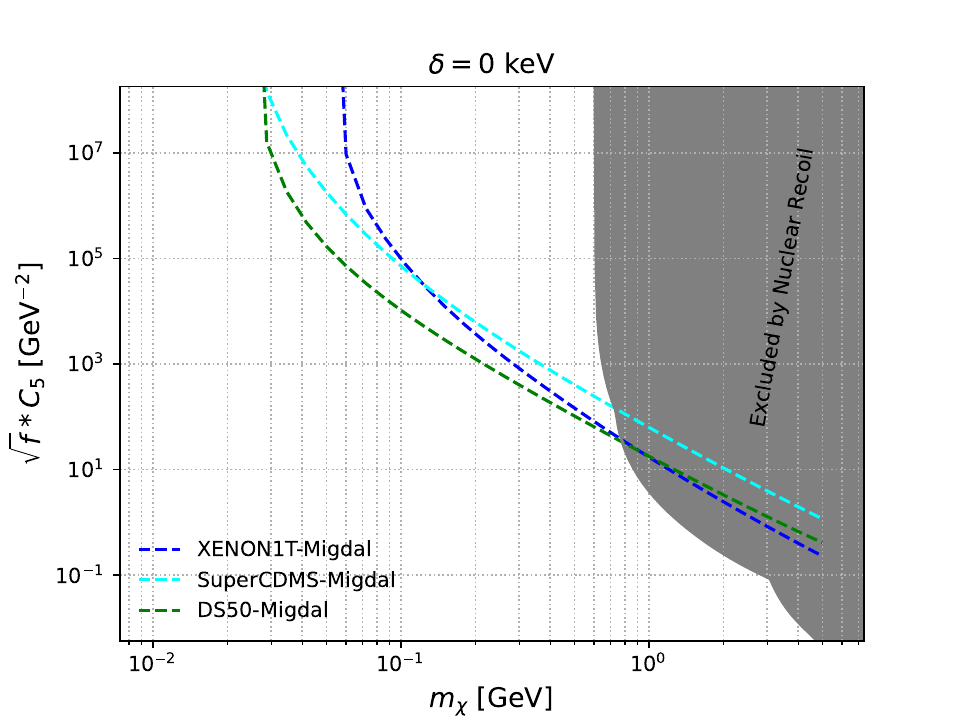}\\
    \includegraphics[width=0.44\textwidth]{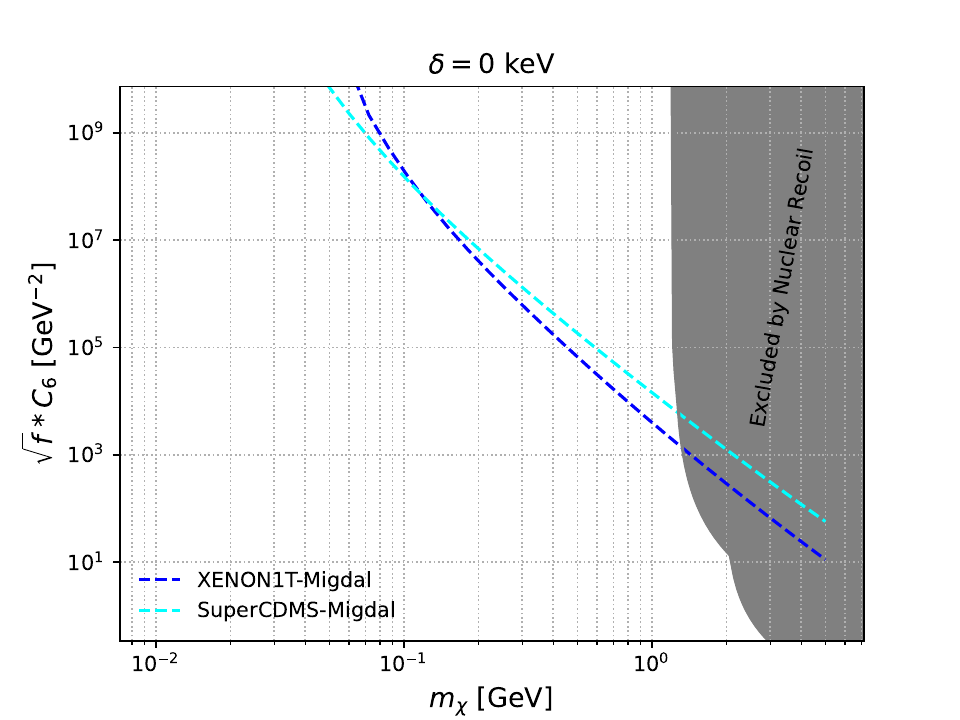}
    \includegraphics[width=0.44\textwidth]{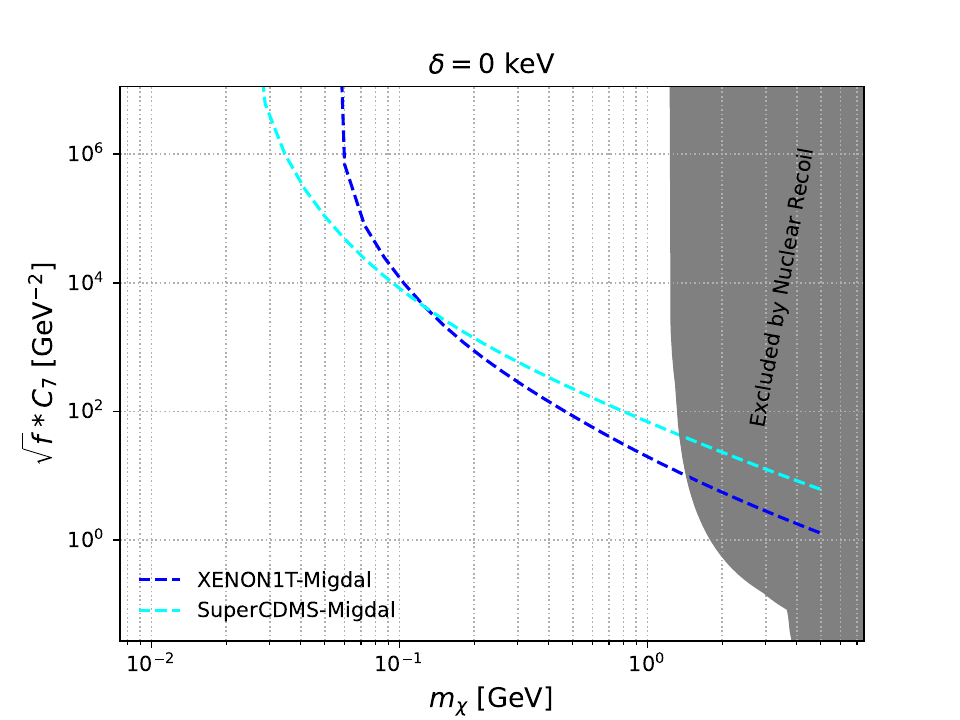}
\caption{Migdal exclusion limit on the combination $\sqrt{f}*c_j$ (with $f$ the fraction of DM local density provided by the $\chi$ particle) for the XENON1T (blue-dashed), SuperCDMS (cyan-dashed), and DS50 (green-dashed) experiments and for the $O_{1,3,4 - 7}$ interactions, considering elastic scattering, $\delta=0$ keV. The grey shaded region represents the combined exclusion limit from experiments searching for nucleus scattering events.}
\label{fig:xenon_migal_exclusions_elastic_1}
\end{center}
\end{figure*}

\begin{figure*}
\begin{center}
    \includegraphics[width=0.44\textwidth]{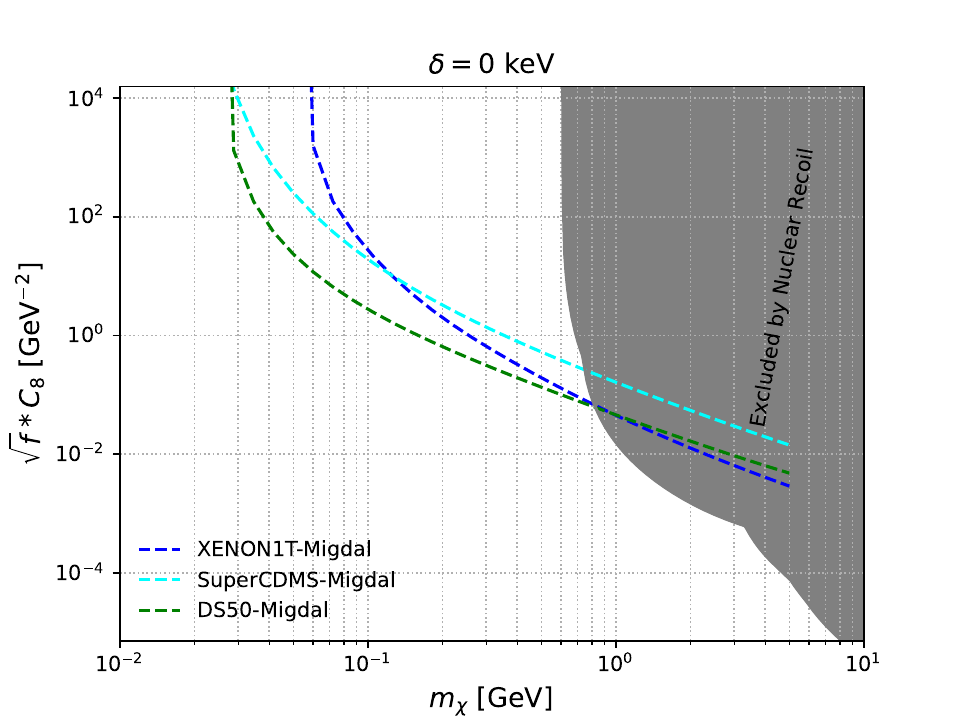}
    \includegraphics[width=0.44\textwidth]{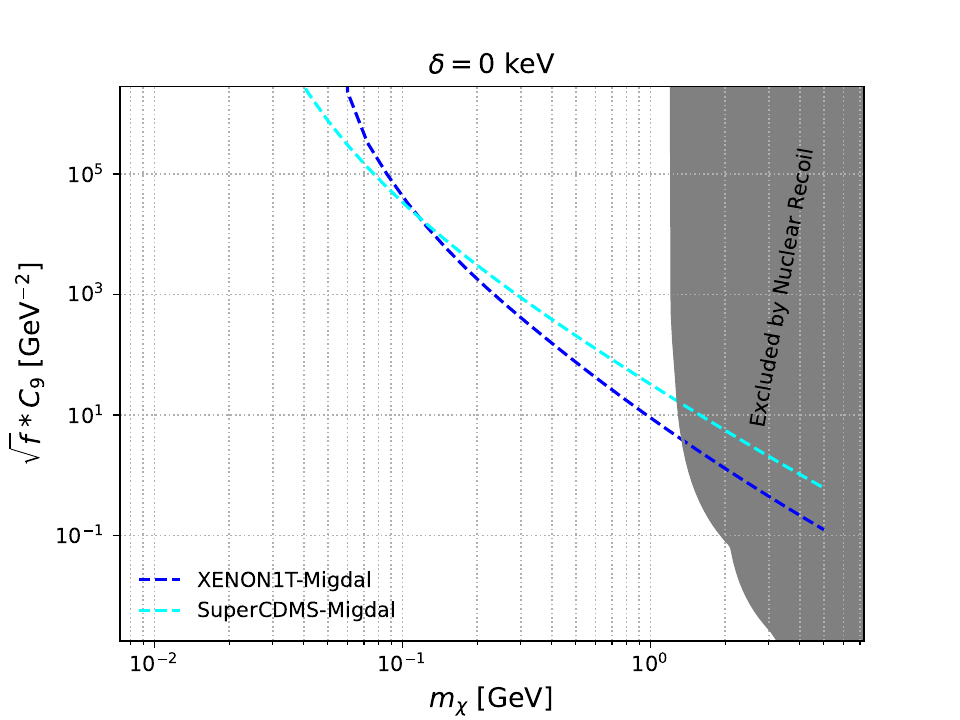}\\
    \includegraphics[width=0.44\textwidth]{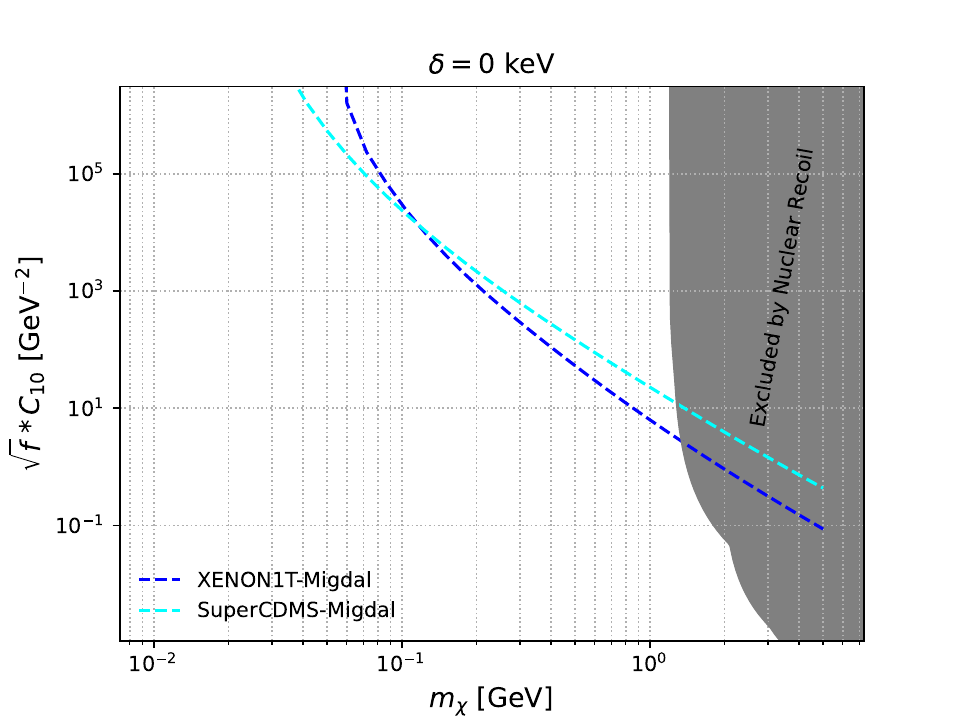}
    \includegraphics[width=0.44\textwidth]{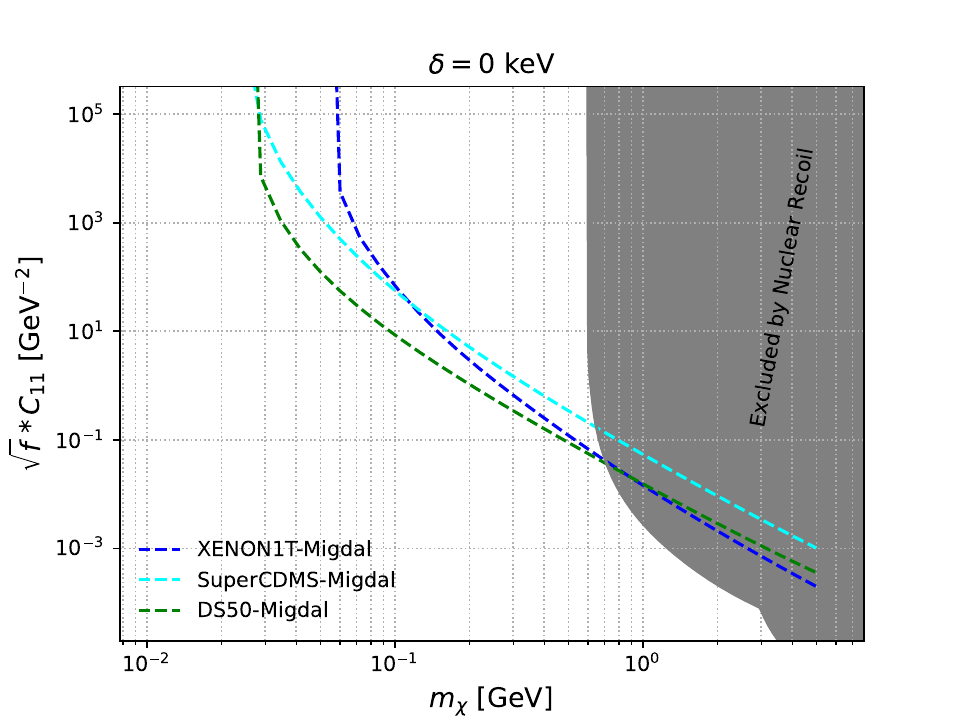}\\
    \includegraphics[width=0.44\textwidth]{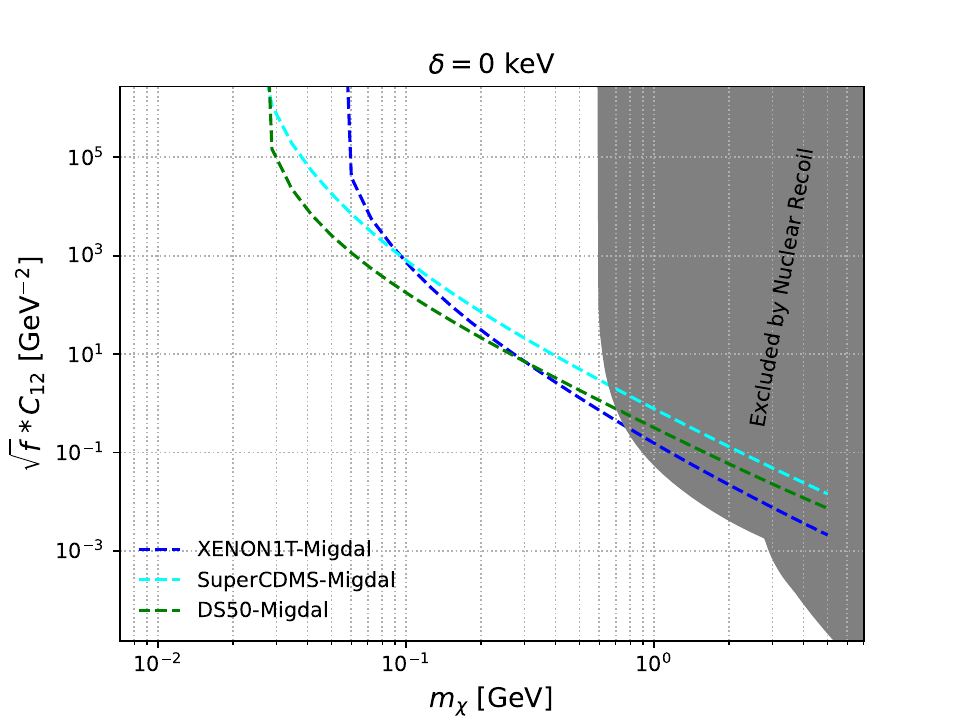}
    \includegraphics[width=0.44\textwidth]{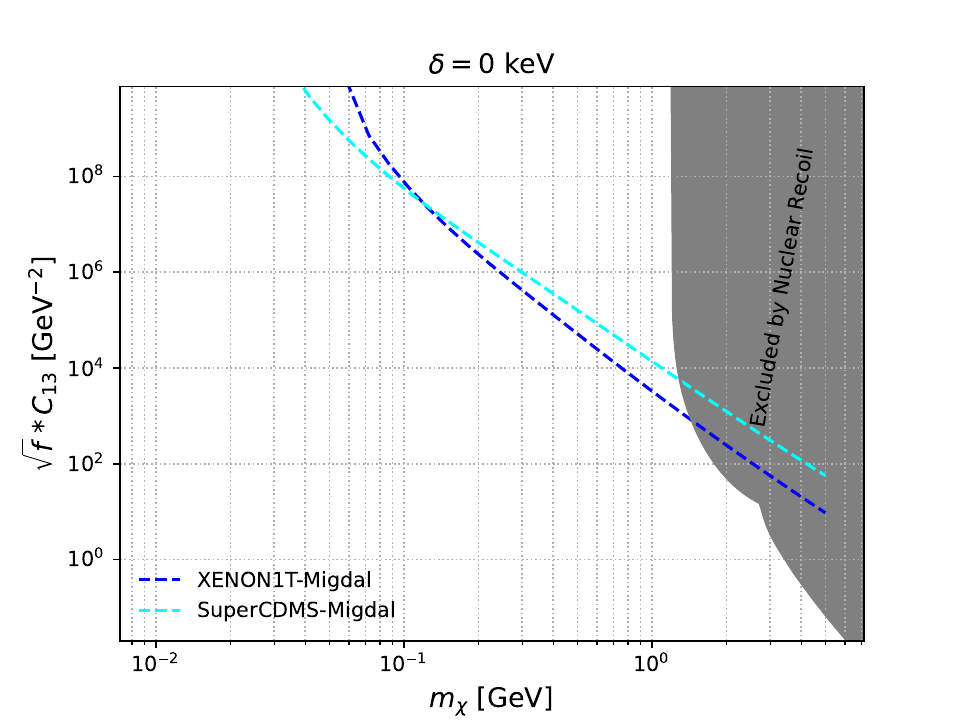}
\caption{ Same as Fig.~\ref{fig:xenon_migal_exclusions_elastic_1} for the
$O_{8,9,10 - 13}$ interactions.}
\label{fig:xenon_migal_exclusions_elastic_2}
\end{center}
\end{figure*}

\begin{figure*}
\begin{center}
    \includegraphics[width=0.44\textwidth]{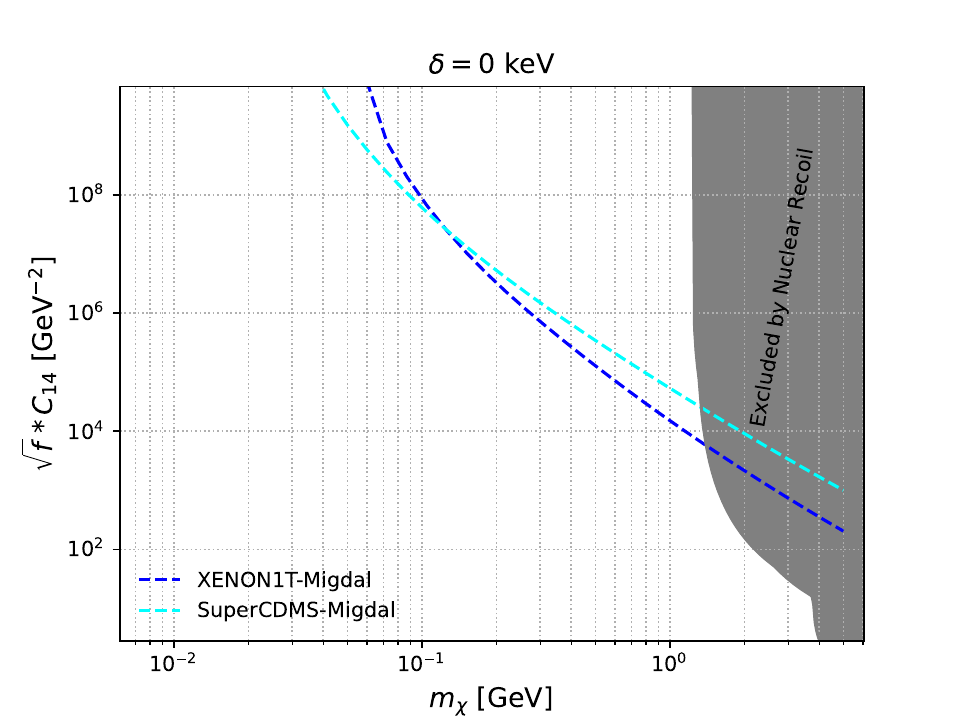}
    \includegraphics[width=0.44\textwidth]{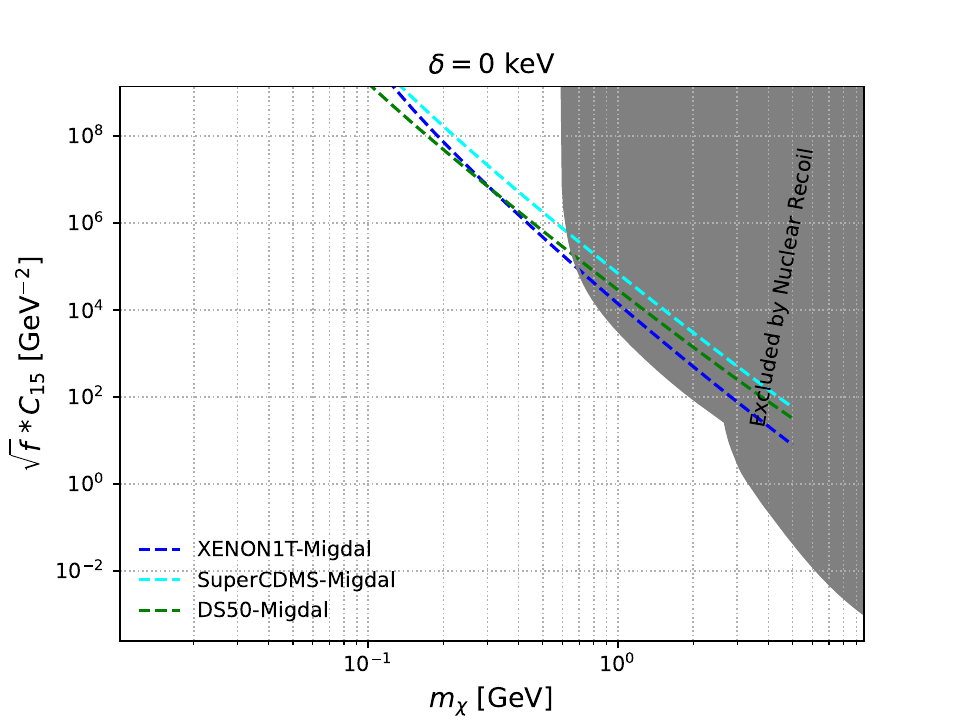}
\caption{Same as Fig.~\ref{fig:xenon_migal_exclusions_elastic_1} for 
$O_{14}$ and ${O_{15}}$ interactions.}
\label{fig:xenon_migal_exclusions_elastic_3}
\end{center}
\end{figure*}

\begin{figure*}
\begin{center}
    \includegraphics[width=0.44\textwidth]{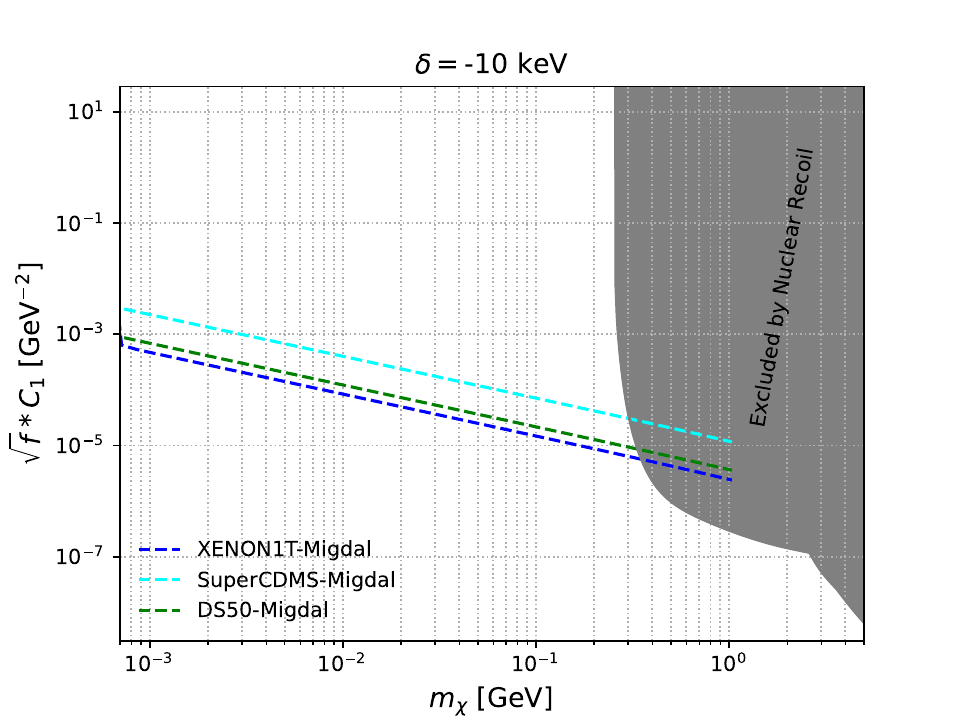}
    \includegraphics[width=0.44\textwidth]{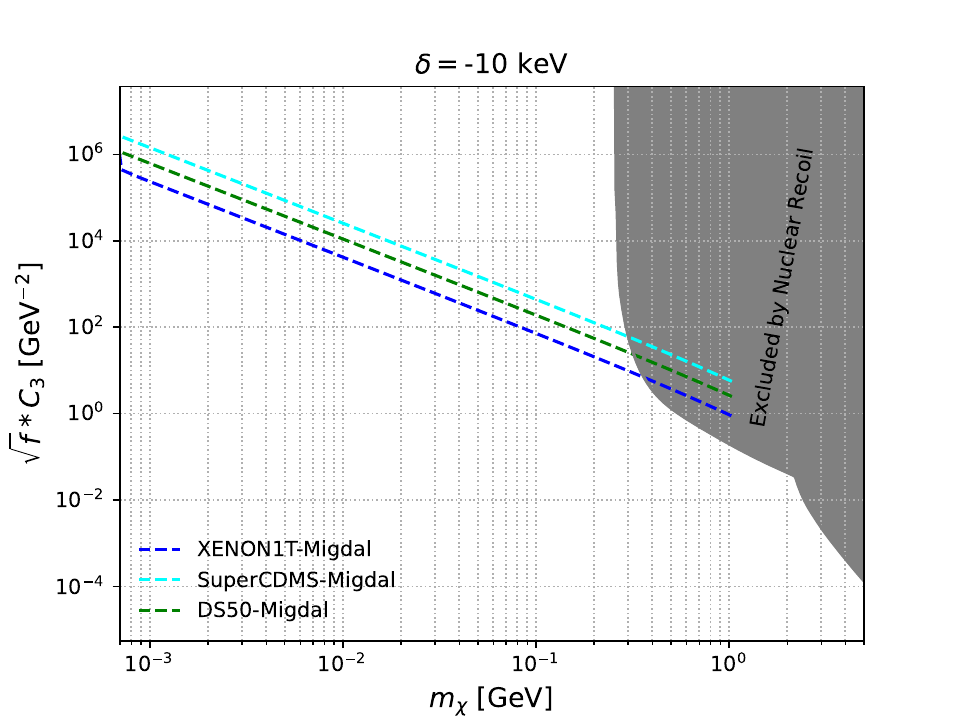}\\
    \includegraphics[width=0.44\textwidth]{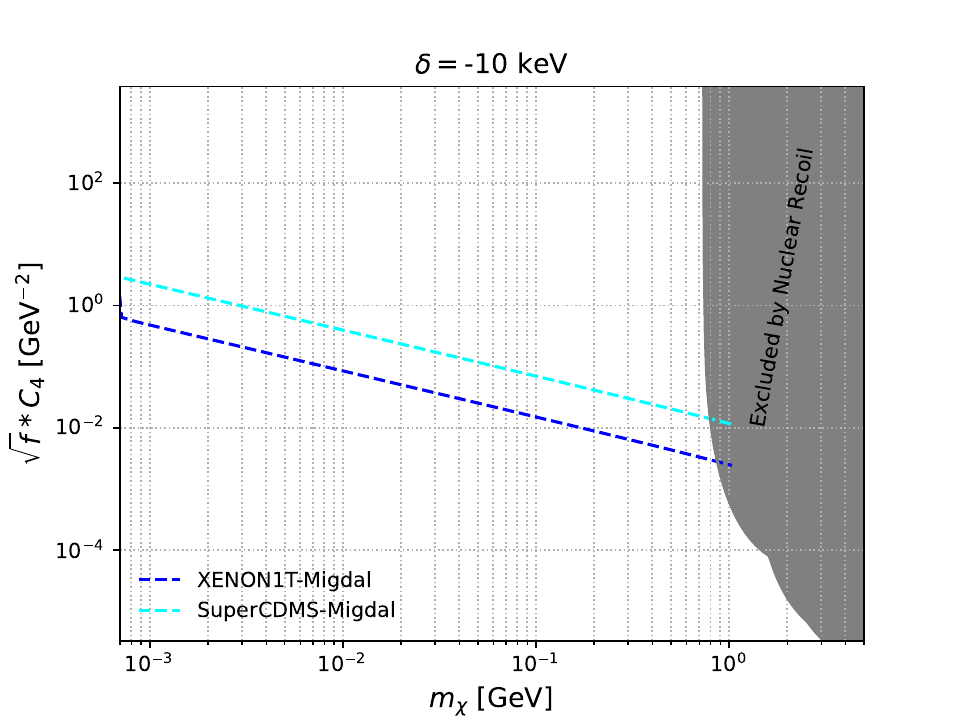}
    \includegraphics[width=0.44\textwidth]{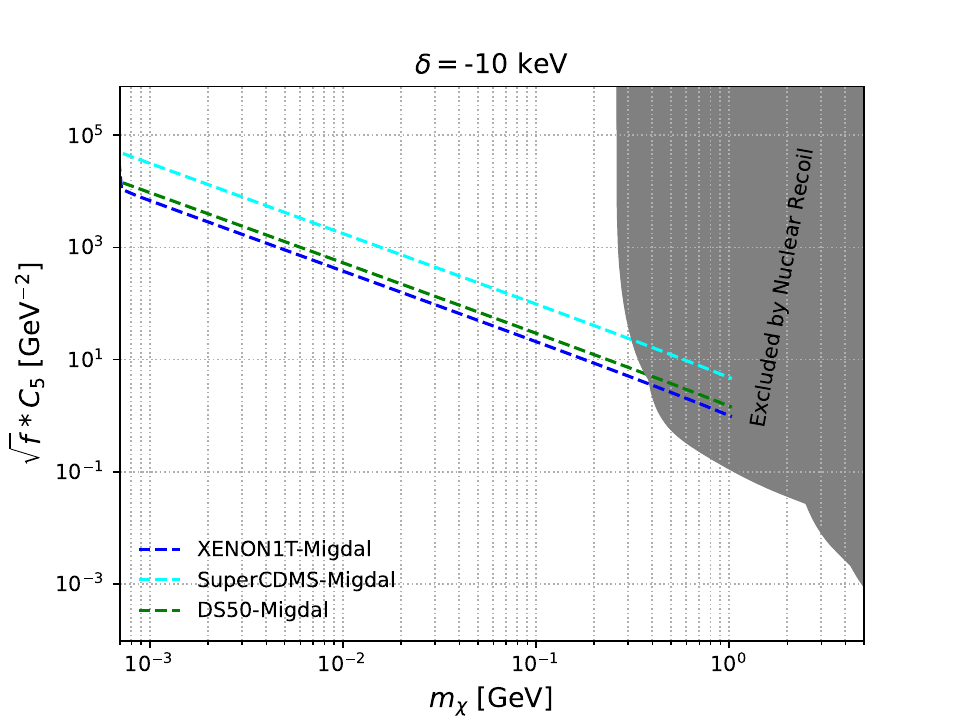}\\
    \includegraphics[width=0.44\textwidth]{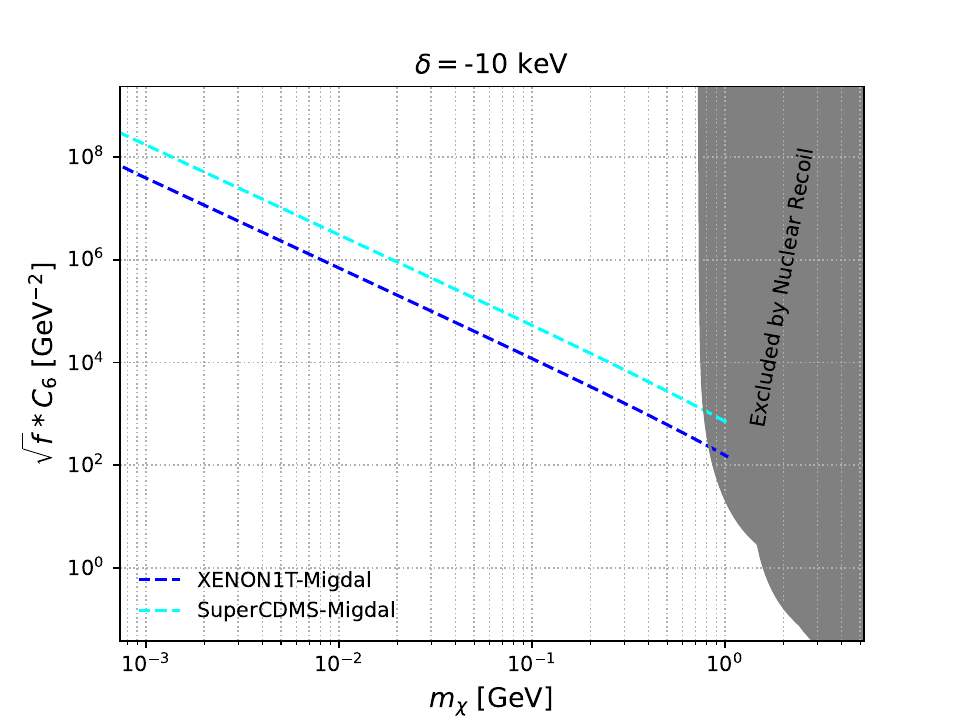}
    \includegraphics[width=0.44\textwidth]{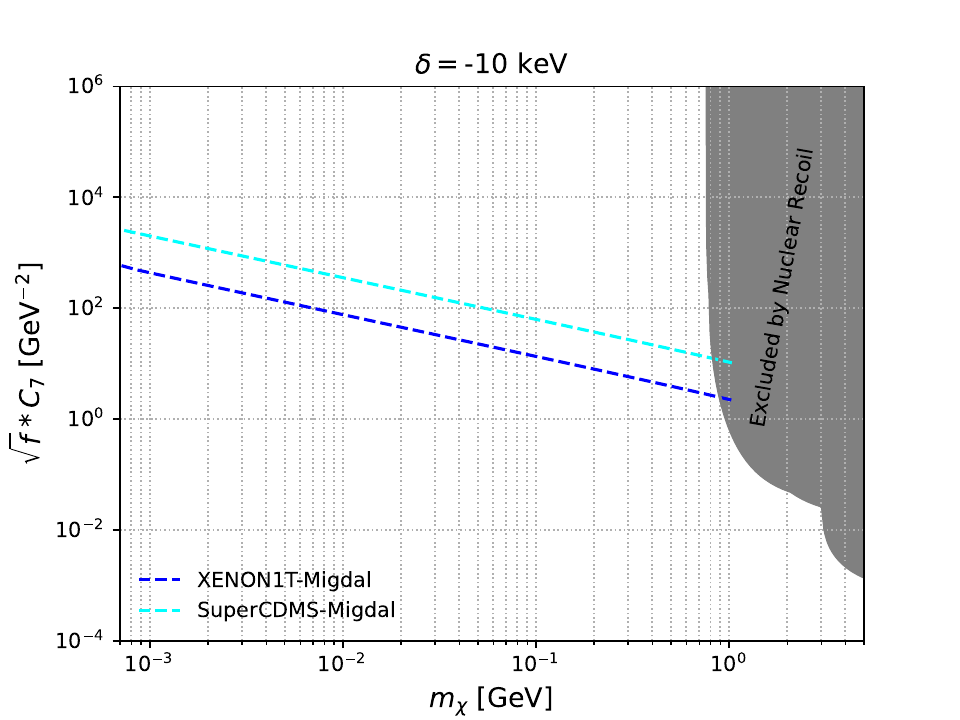}
\caption{Migdal exclusion limit for XENON1T (blue), SuperCDMS (cyan), and DS50 (green) experiments for $O_{1,3,4 - 7}$ interactions, considering exothermic scattering, $\delta=-10$ keV. Combined exclusion limit from exothermic DM--nucleus scattering in direct detection experiments is presented as a grey shaded region.}
\label{fig:xenon_migal_exclusions_exothermic_1}
\end{center}
\end{figure*}

\begin{figure*}
\begin{center}
    \includegraphics[width=0.44\textwidth]{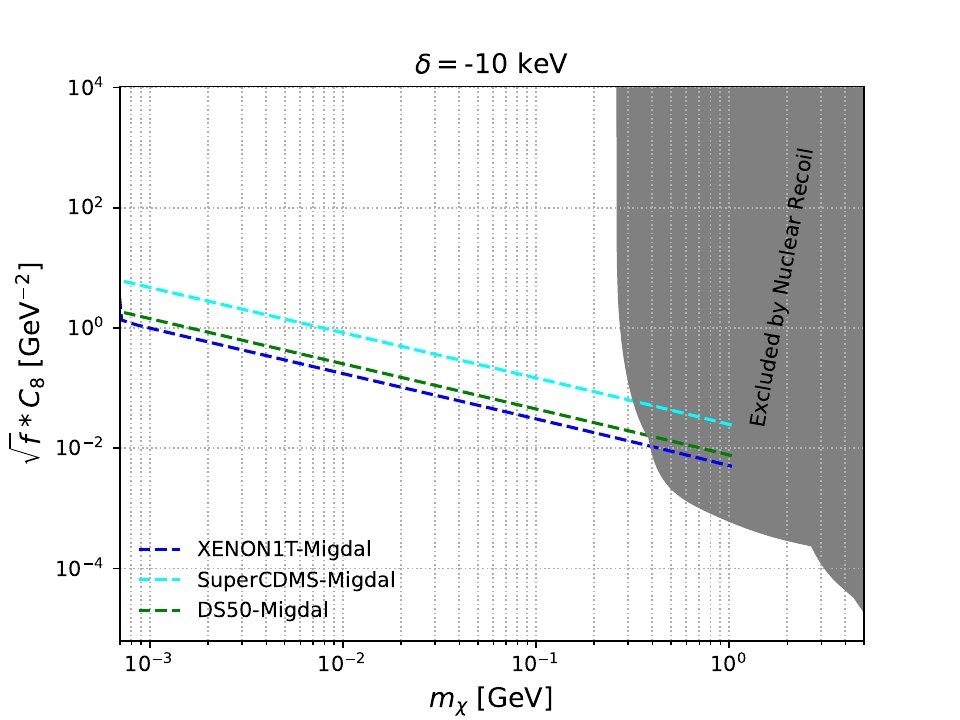}
    \includegraphics[width=0.44\textwidth]{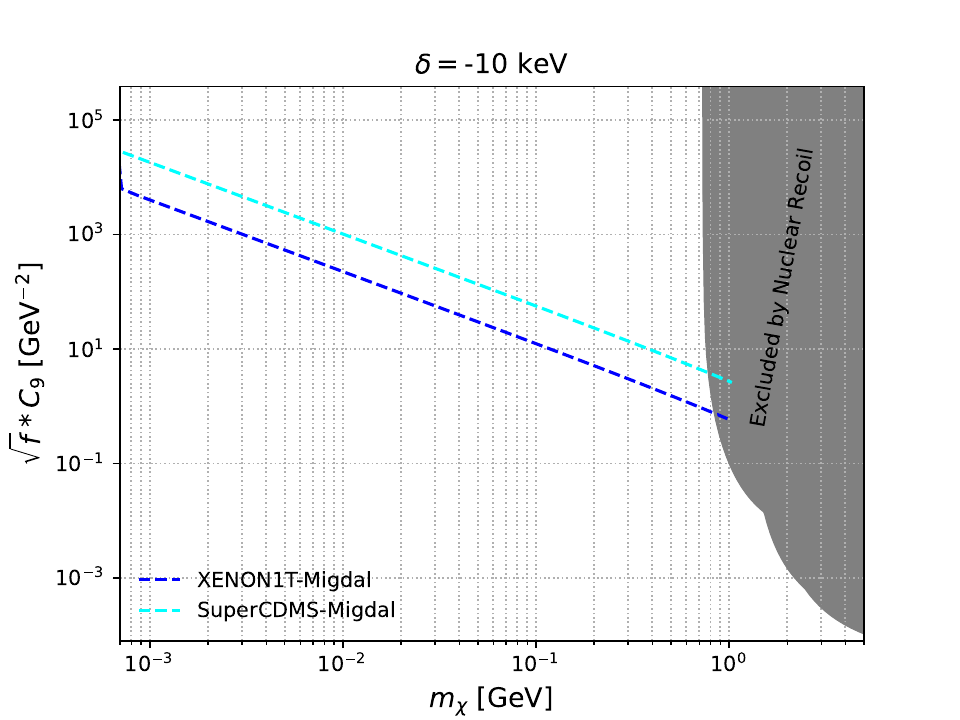}\\
    \includegraphics[width=0.44\textwidth]{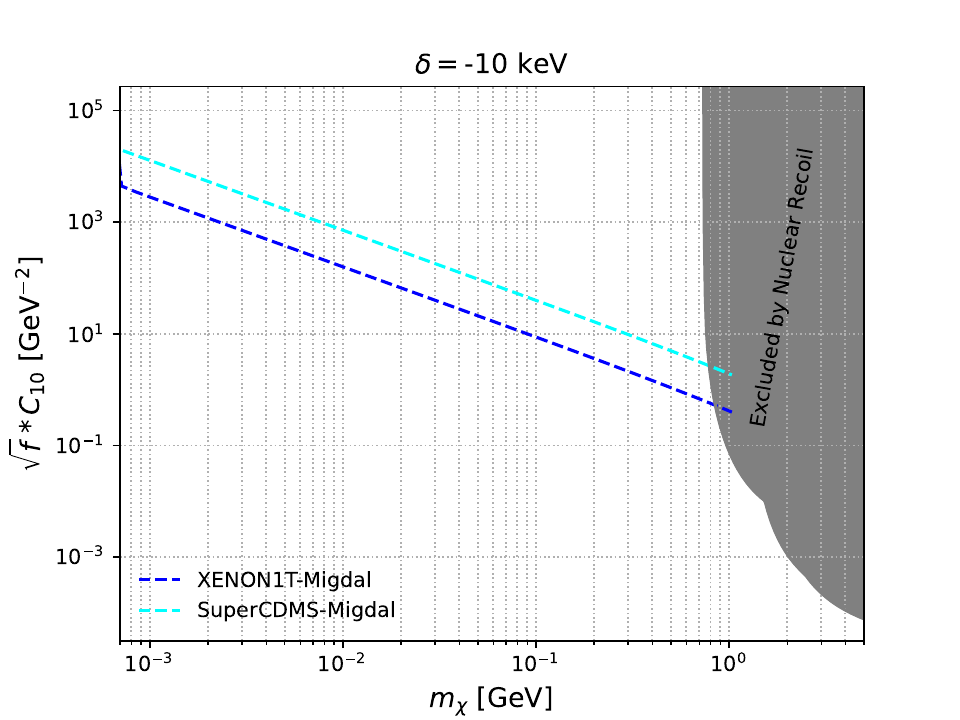}
    \includegraphics[width=0.44\textwidth]{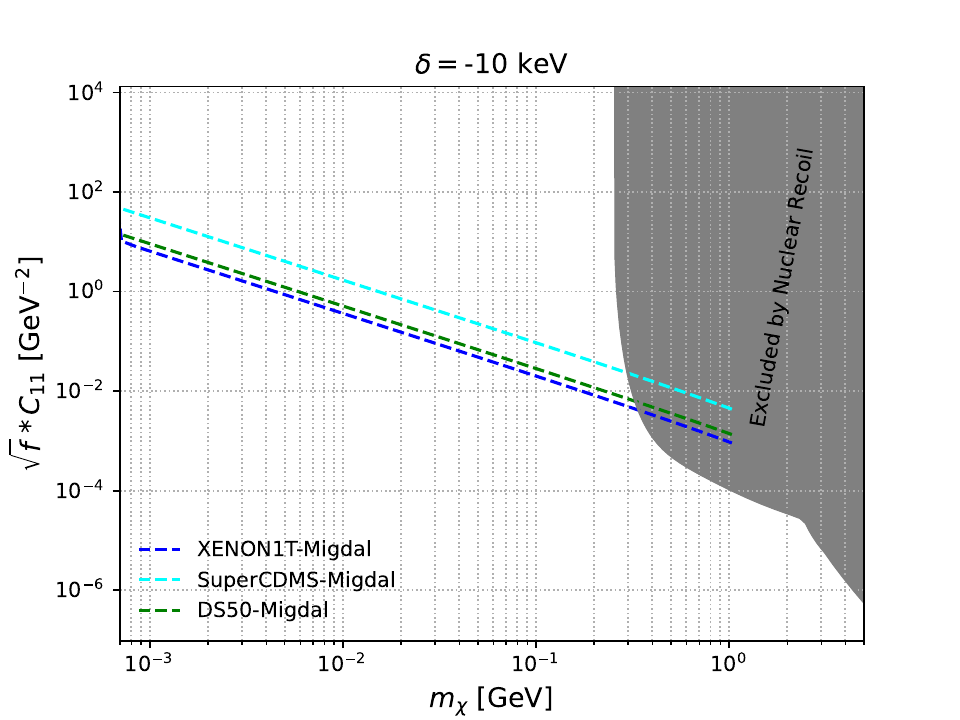}\\
    \includegraphics[width=0.44\textwidth]{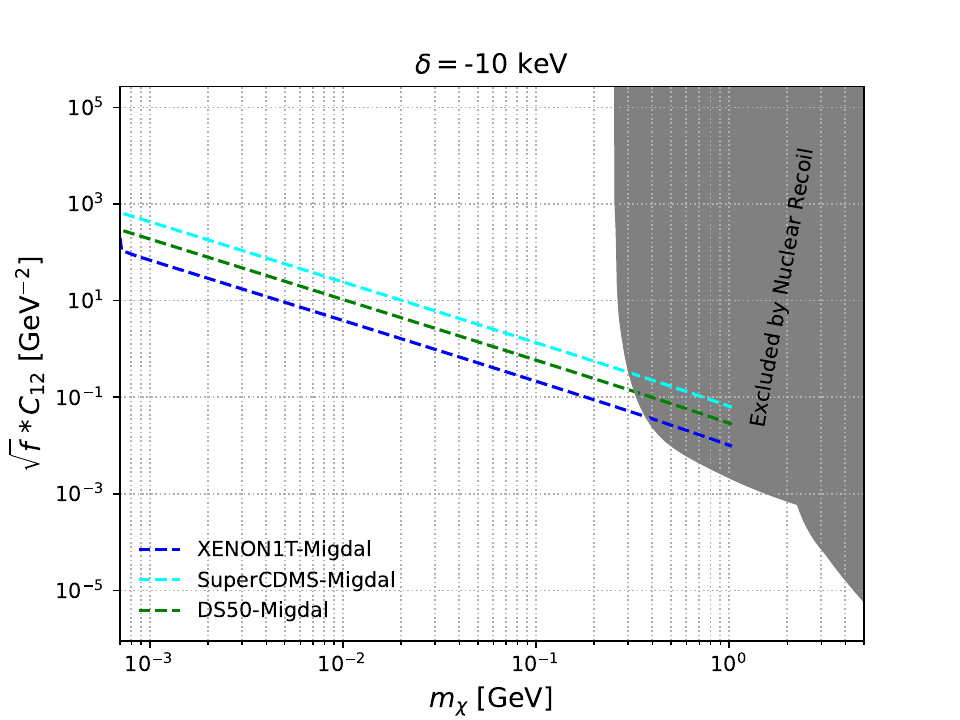}
    \includegraphics[width=0.44\textwidth]{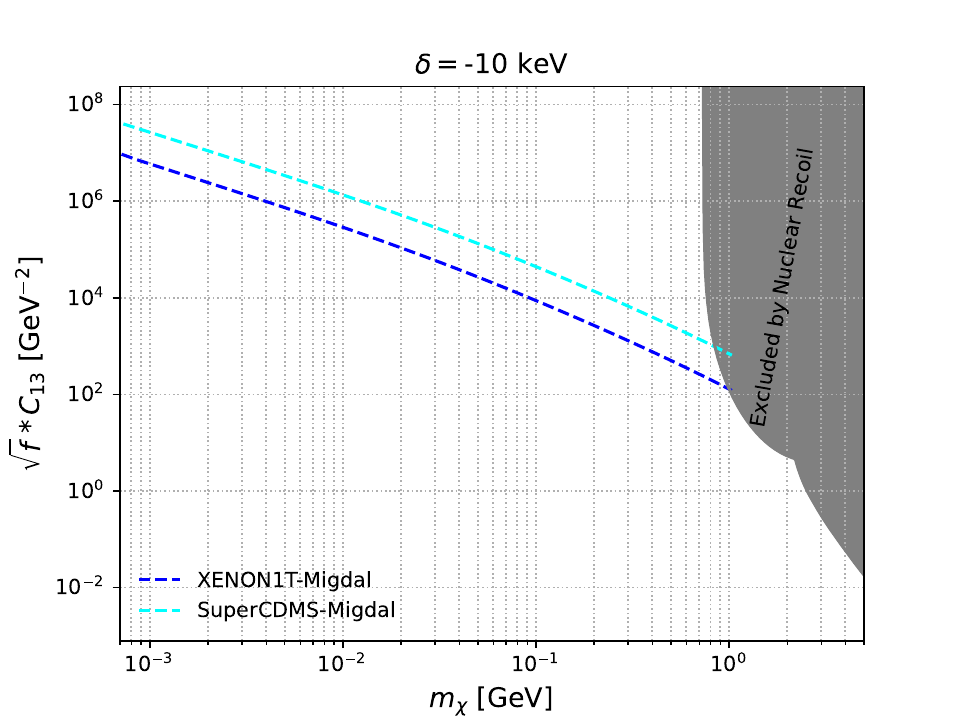}
\caption{Same as Fig.~\ref{fig:xenon_migal_exclusions_exothermic_1} for 
$O_{8,9,10 - 13}$ interactions}
\label{fig:xenon_migal_exclusions_exothermic_2}
\end{center}
\end{figure*}

\begin{figure*}
\begin{center}
    \includegraphics[width=0.44\textwidth]{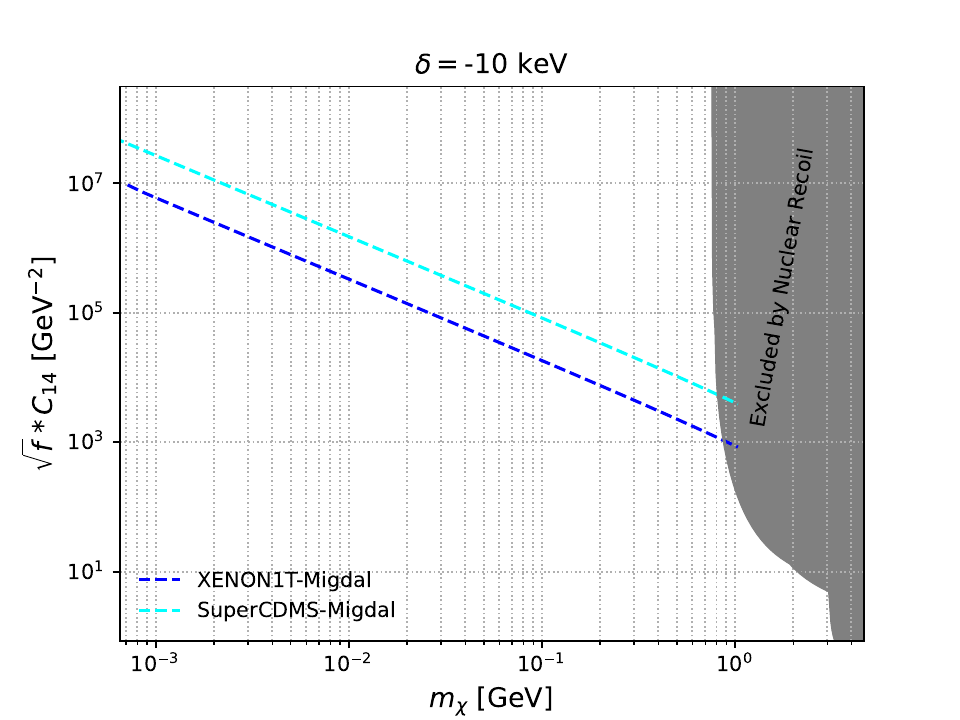}
    \includegraphics[width=0.44\textwidth]{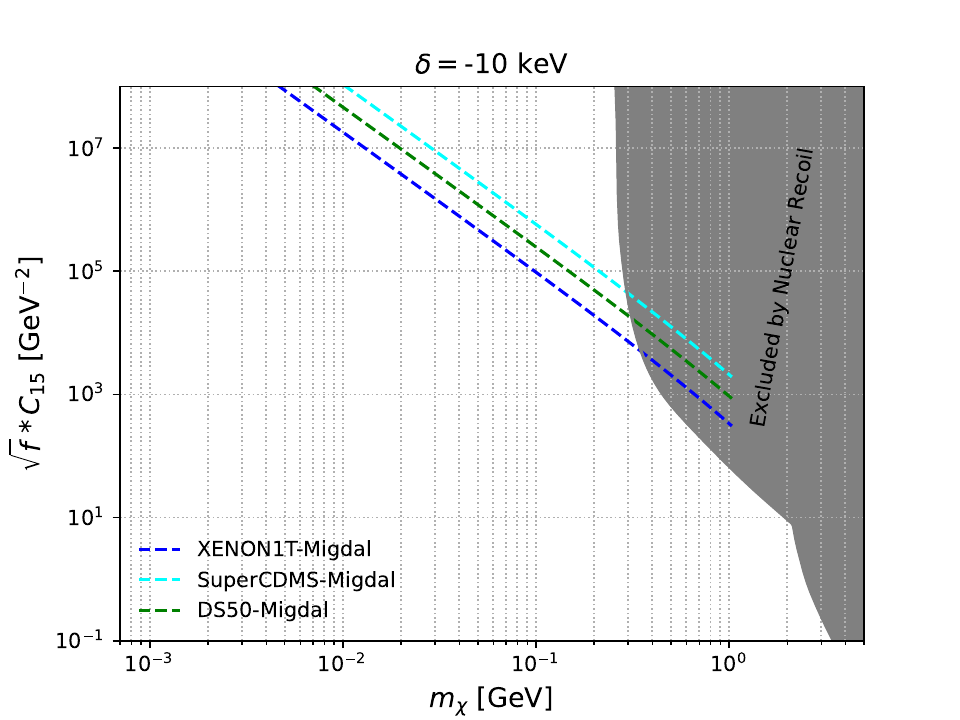}
\caption{Same as Fig.~\ref{fig:xenon_migal_exclusions_exothermic_1} for 
$O_{14}$ and $O_{15}$ interactions}
\label{fig:xenon_migal_exclusions_exothermic_3}
\end{center}
\end{figure*}

\begin{figure*}
\begin{center}
    \includegraphics[width=0.45\textwidth]{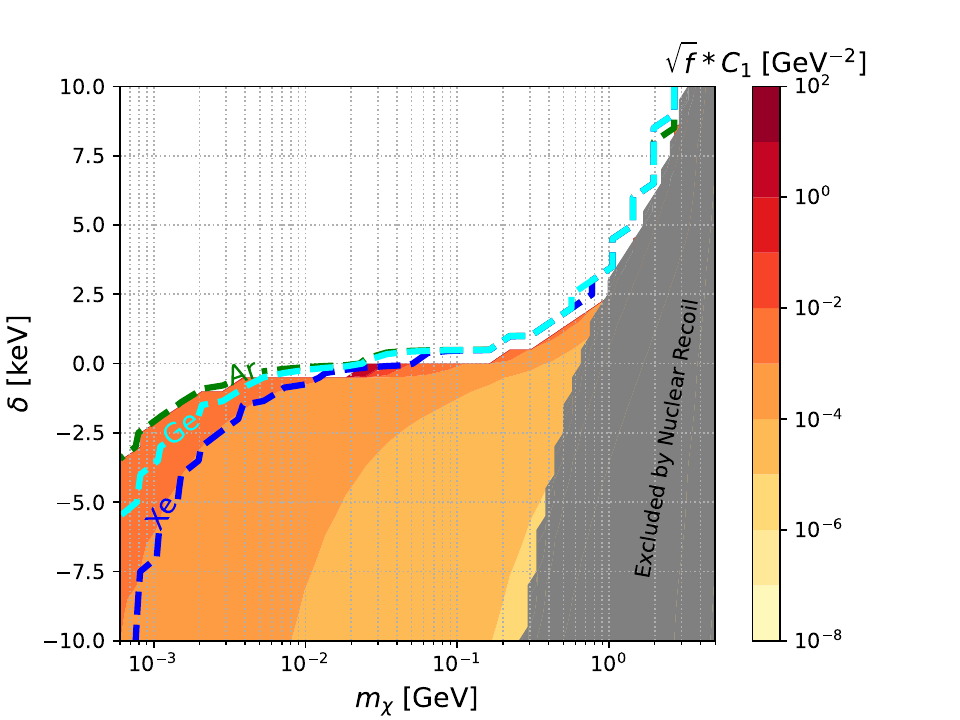}
    \includegraphics[width=0.45\textwidth]{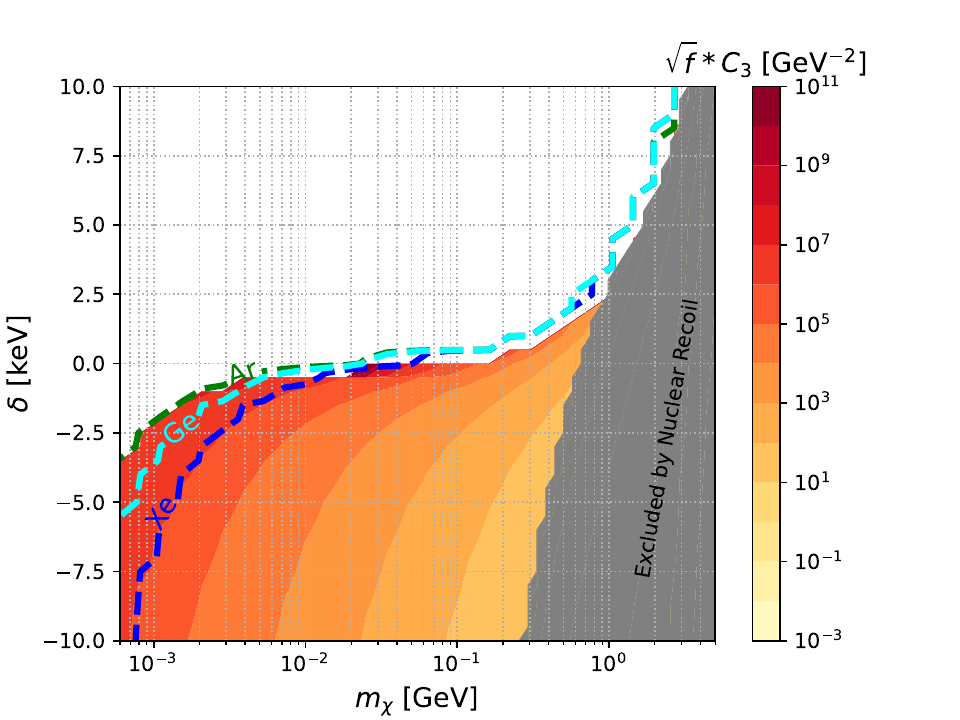}\\
    \includegraphics[width=0.45\textwidth]{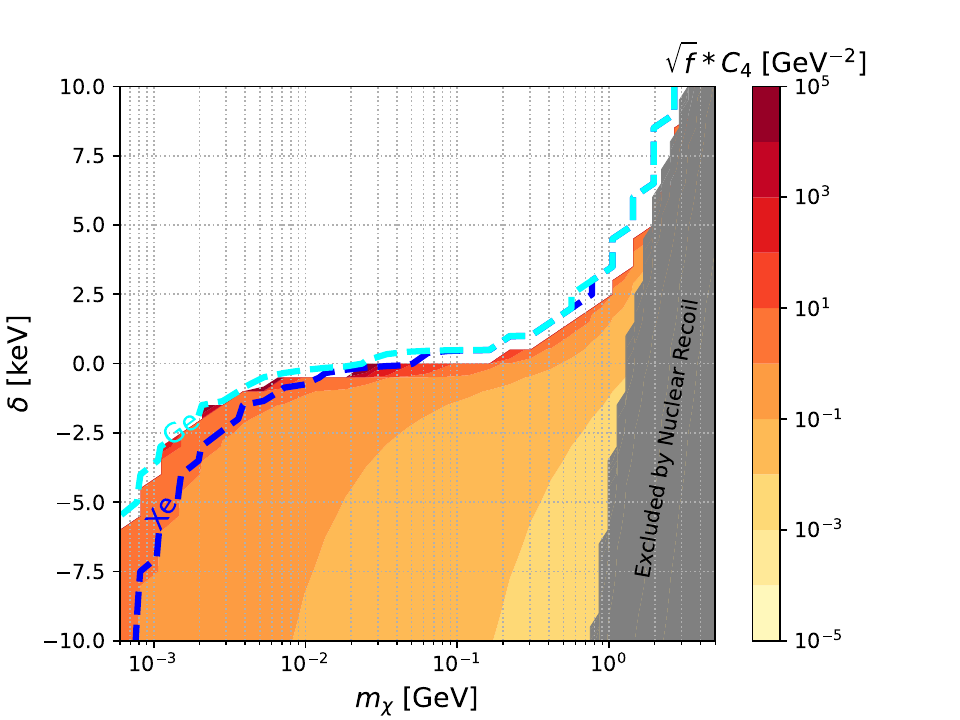}
    \includegraphics[width=0.45\textwidth]{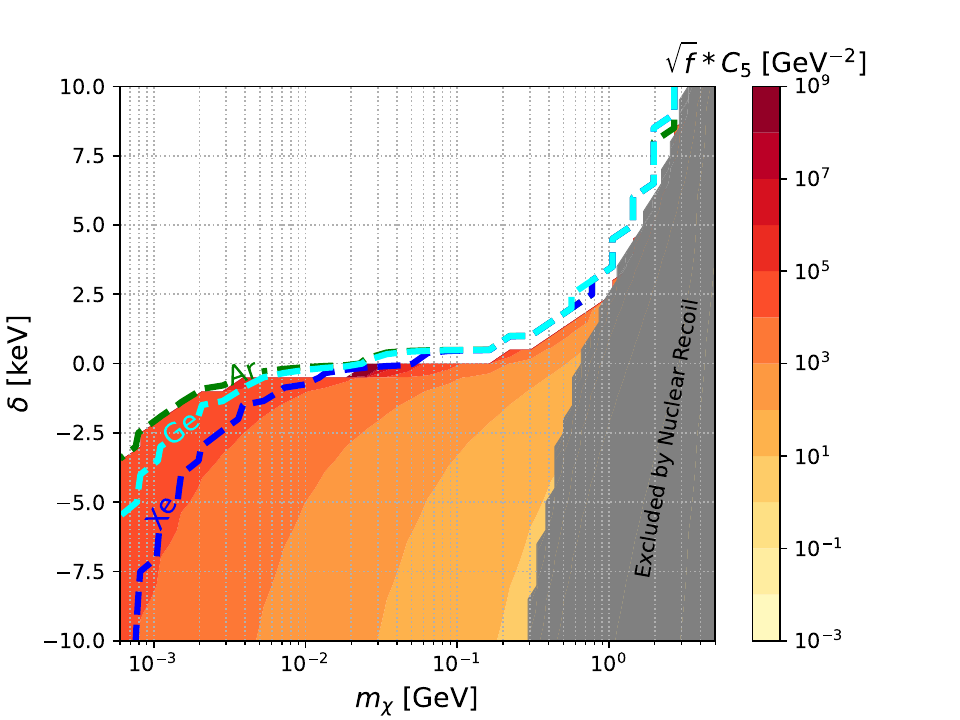}\\
    \includegraphics[width=0.45\textwidth]{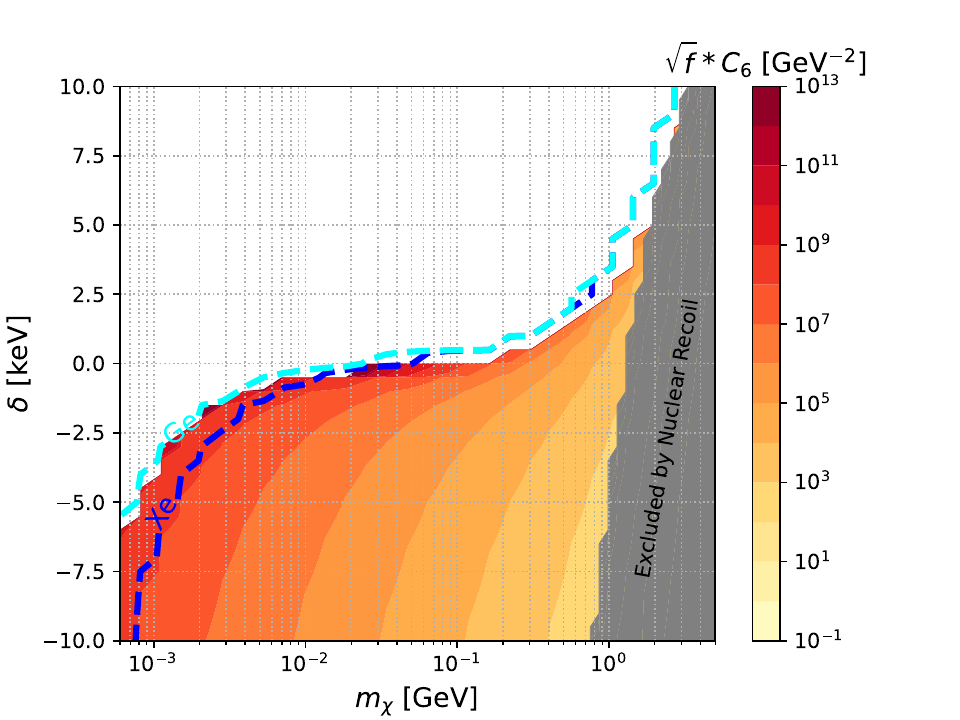}
    \includegraphics[width=0.45\textwidth]{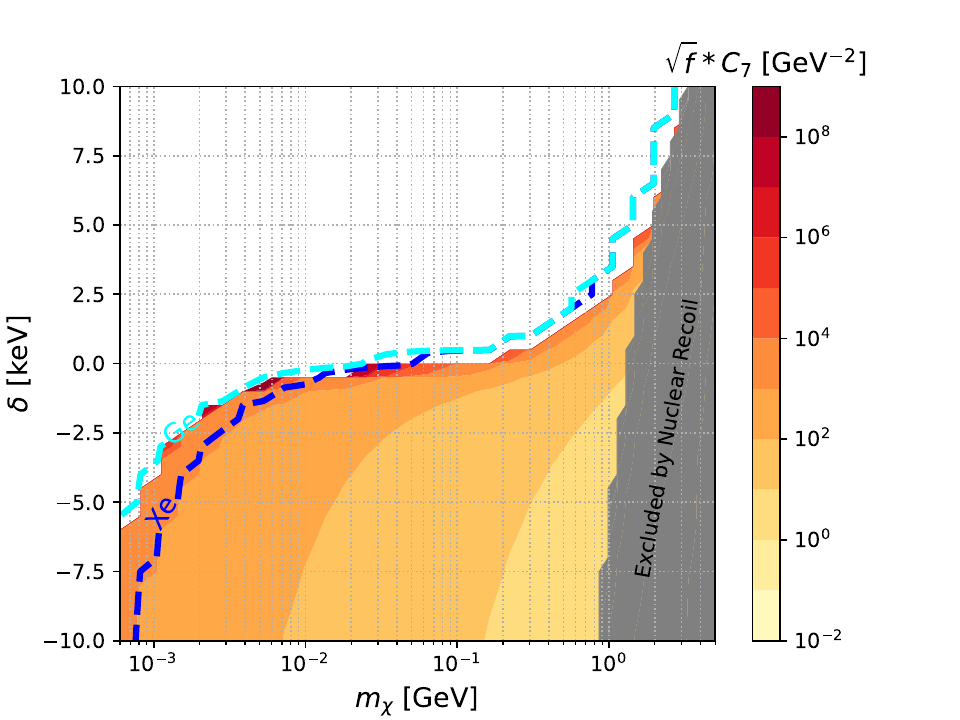}
\caption{Combined exclusion plots for $O_{1,2,3 - 7}$ interactions are presented in the $m_\chi-\delta$ plane. The color gradient delineates regions excluded by Migdal data from XENON1T, SuperCDMS, and DS50 experiments, while the shaded grey region represents data from direct detection experiments included in our analysis.
Additionally, dashed lines in green, cyan, and blue denote $E^{max}_R<50$ meV for Ar, Ge, and Xe targets, respectively, obtained through the impulse approximation.}
\label{fig:exclusion_contour_1}
\end{center}
\end{figure*}

\begin{figure*}
\begin{center}
    \includegraphics[width=0.45\textwidth]{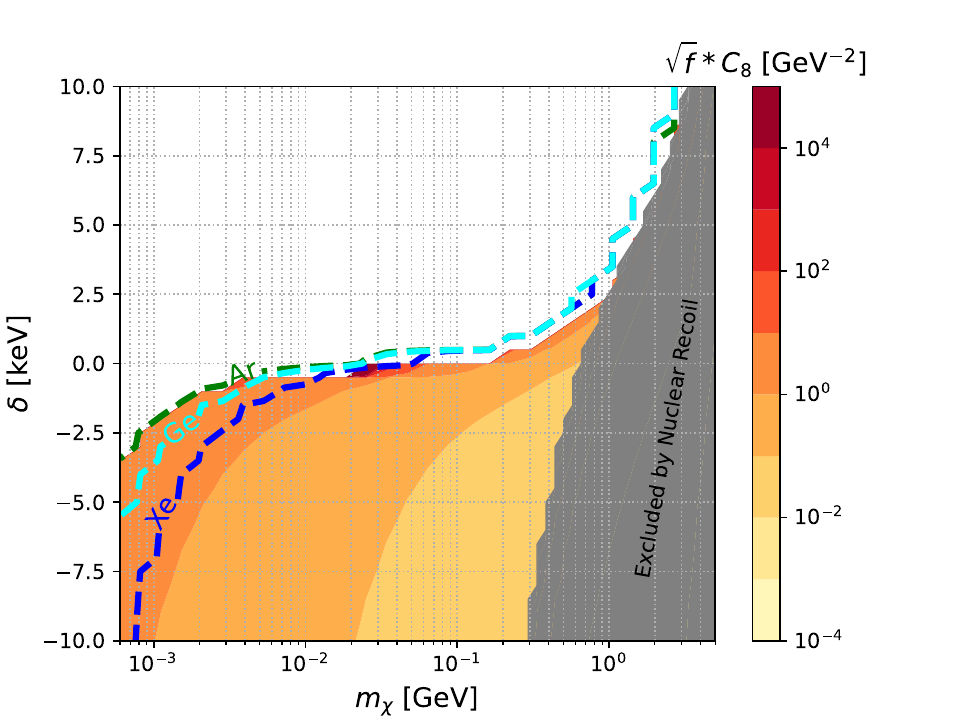}
    \includegraphics[width=0.45\textwidth]{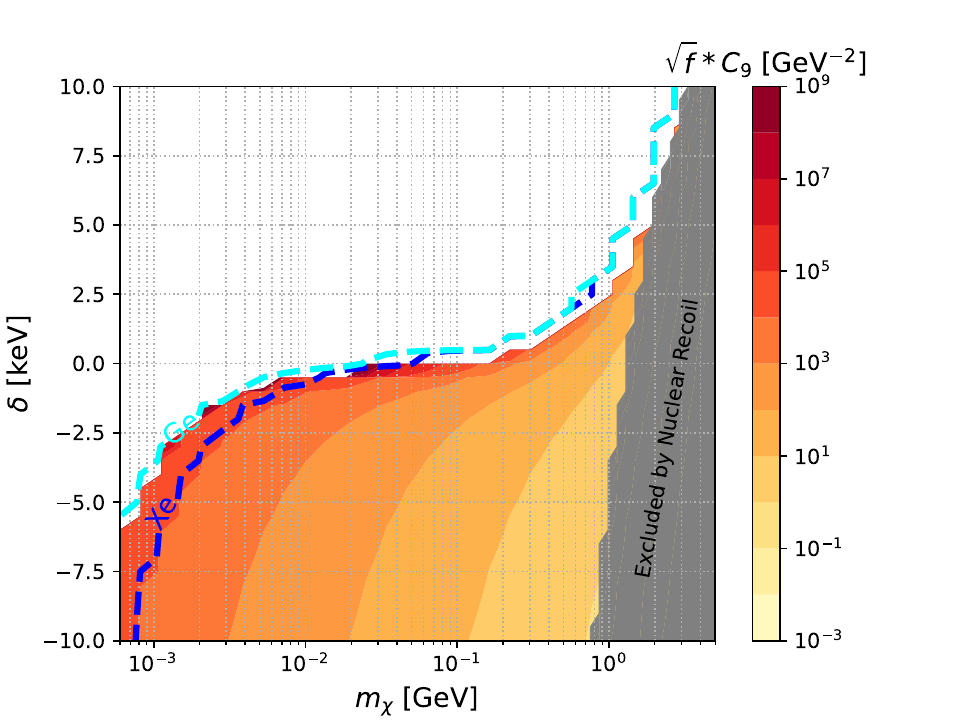}\\
    \includegraphics[width=0.45\textwidth]{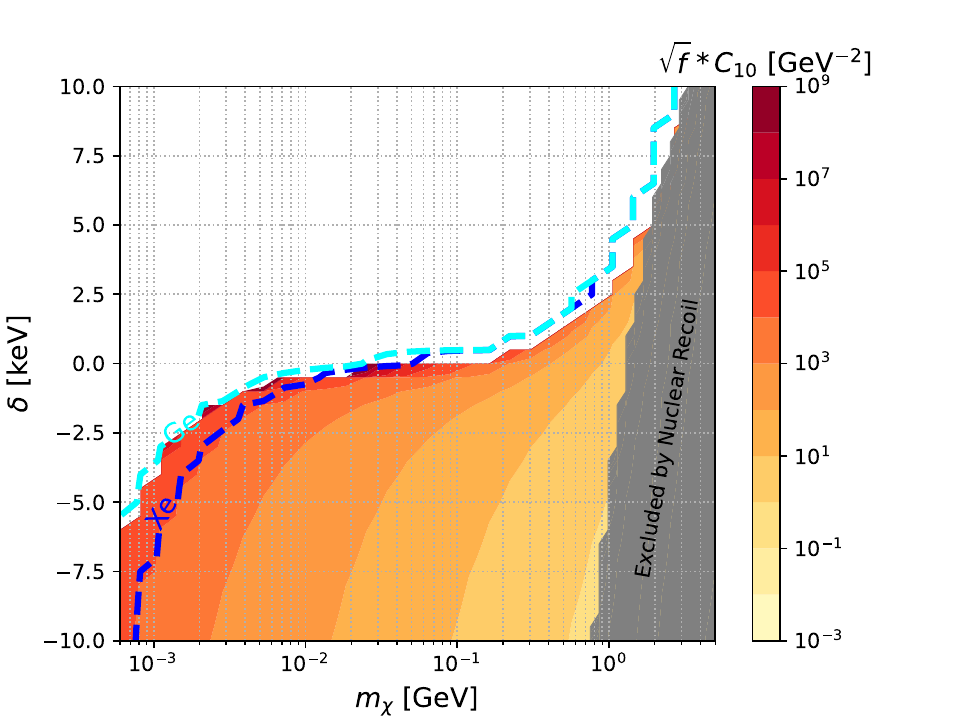}
    \includegraphics[width=0.45\textwidth]{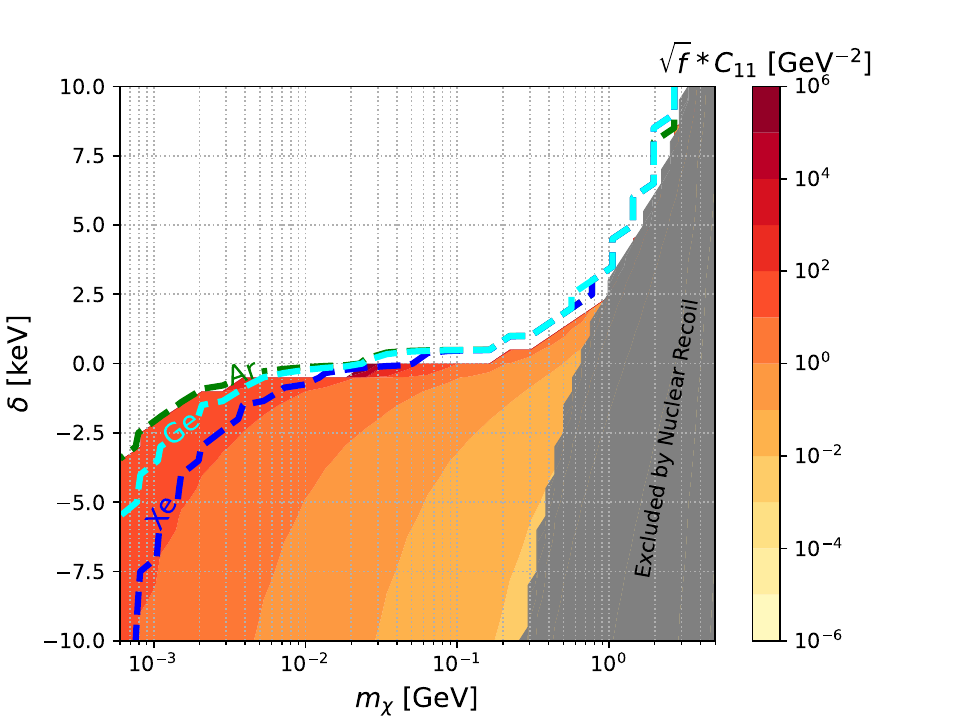}\\
    \includegraphics[width=0.45\textwidth]{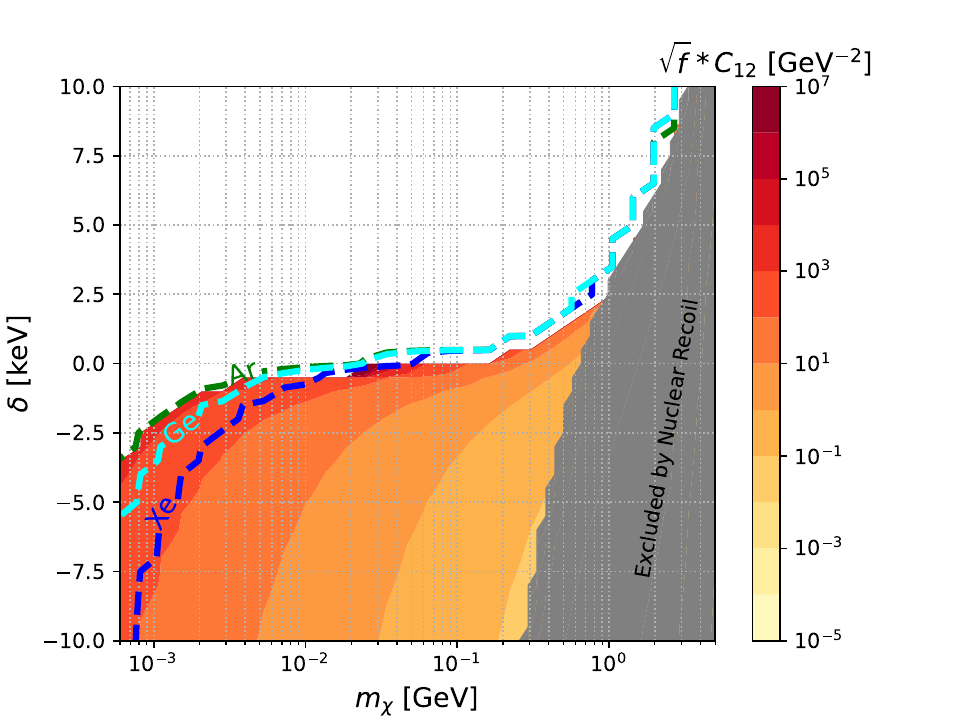}
    \includegraphics[width=0.45\textwidth]{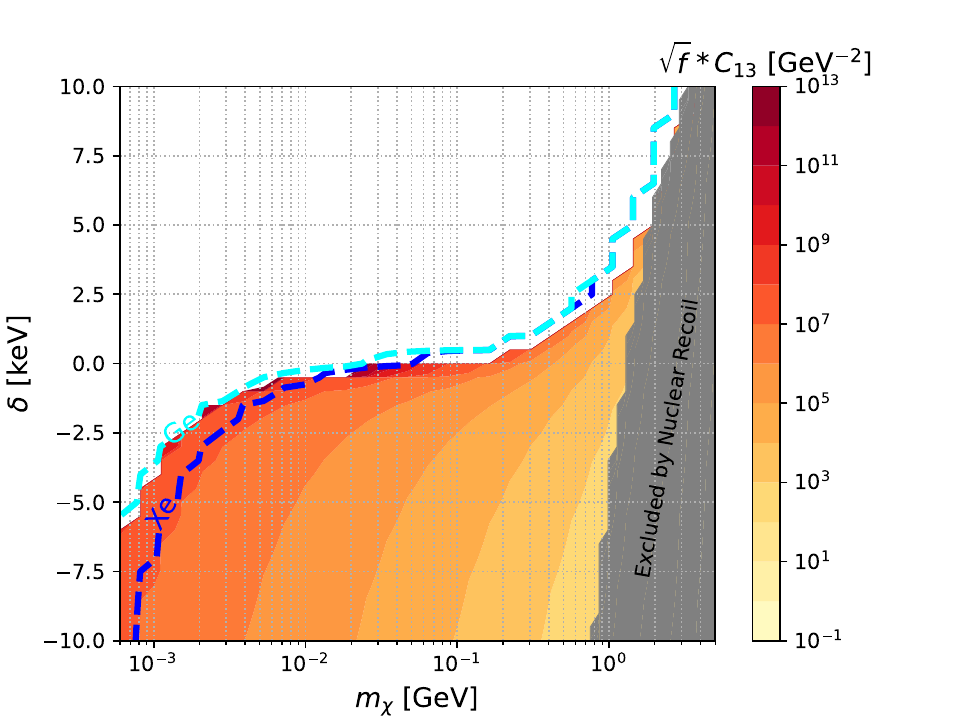}\\
\caption{Same as Fig.~\ref{fig:exclusion_contour_1} for $O_{8,9,10 - 13}$ interactions.}
\label{fig:exclusion_contour_2}
\end{center}
\end{figure*}

\begin{figure*}
\begin{center}
    \includegraphics[width=0.45\textwidth]{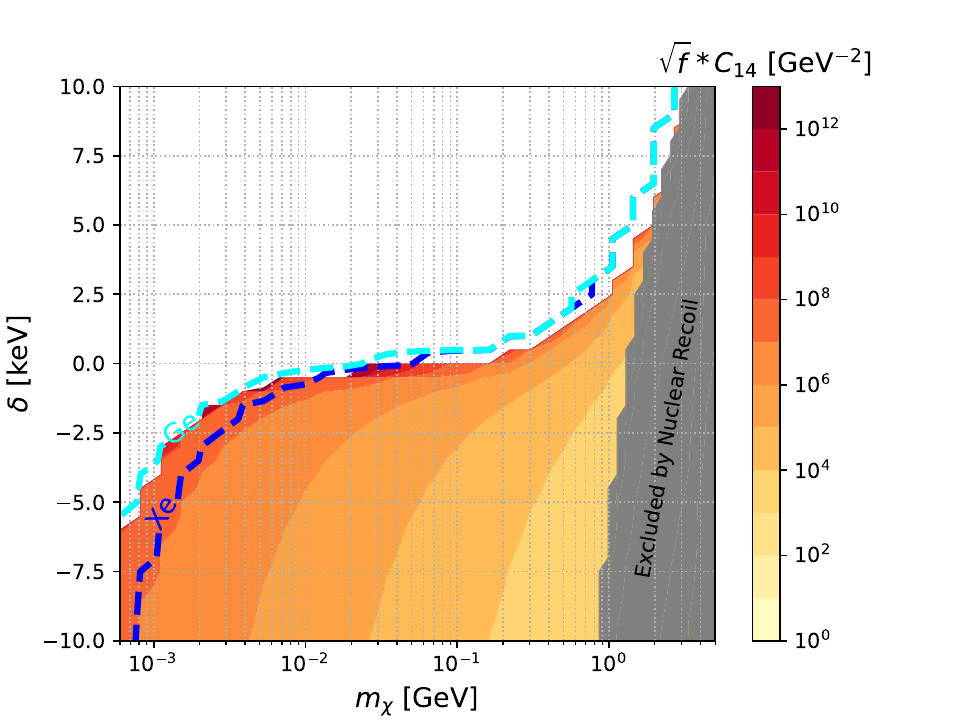}
    \includegraphics[width=0.45\textwidth]{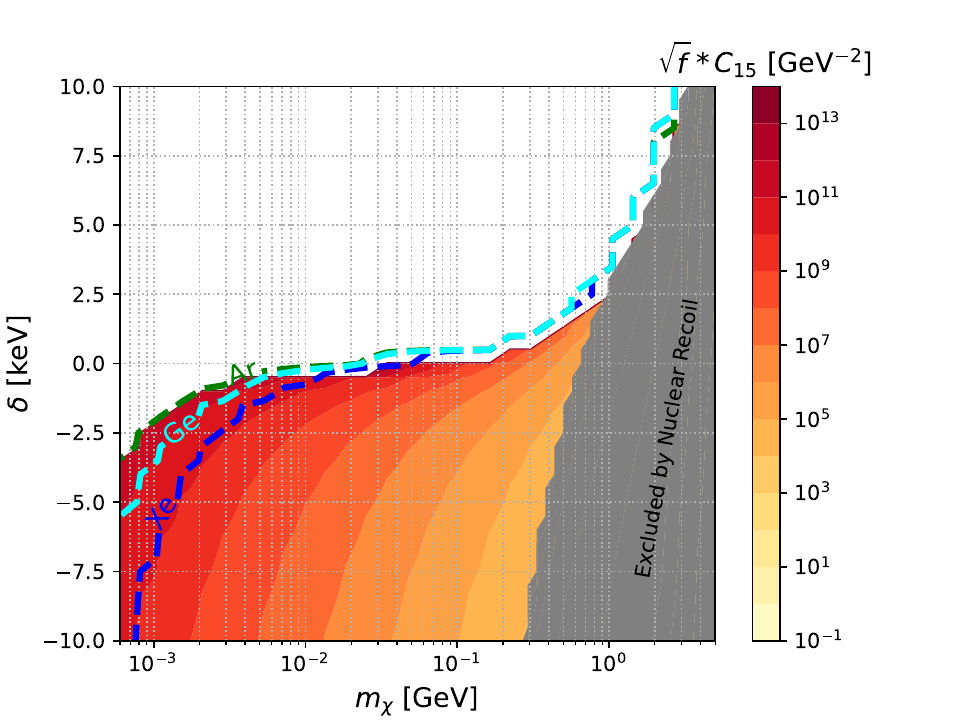}
\caption{Same as Fig.~\ref{fig:exclusion_contour_1} for $O_{14}$ and $O_{15}$ interactions.}
\label{fig:exclusion_contour_3}
\end{center}
\end{figure*}

\begin{figure*}
\begin{center}
    \includegraphics[width=0.45\textwidth]{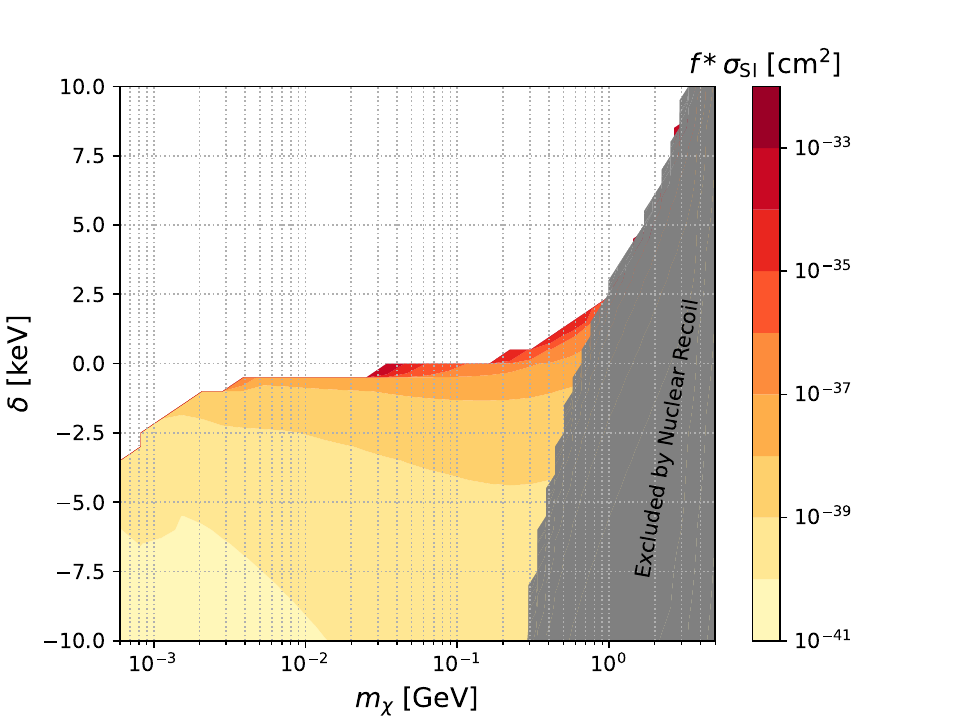}
    \includegraphics[width=0.45\textwidth]{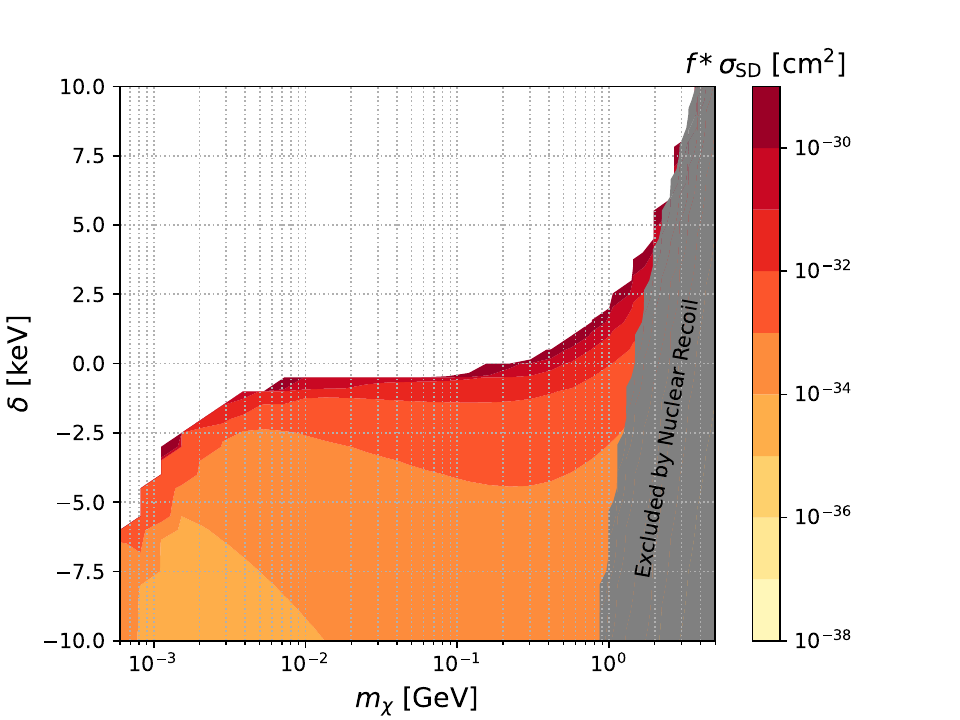}
\caption{Combined exclusion limits for SI and SD interaction in the $m_\chi-\delta$ plane.}
\label{fig:xenon_migal_exclusions_exothermic_cs}
\end{center}
\end{figure*}

The results of our analysis are presented in Figs.~\ref{fig:xenon_migal_exclusions_elastic_1}--\ref{fig:xenon_migal_exclusions_exothermic_cs}. 
In all our plots we assume one non-vanishing coupling at a time in the effective Hamiltonian  of Eq.~(\ref{eq:H}) and an isoscalar interaction, i.e. $c_j^1=0$. In Figs.~\ref{fig:xenon_migal_exclusions_elastic_1}--\ref{fig:xenon_migal_exclusions_elastic_3} the exclusion limits on each effective coupling $c_j$ are shown as a function of $m_\chi$ for $\delta=0$, while in Figs.~\ref{fig:xenon_migal_exclusions_exothermic_1}--\ref{fig:xenon_migal_exclusions_exothermic_3} we show the analogous results for $\delta=-10$ keV.  The exclusion limit on the same couplings are explored in the $m_\chi-\delta$ plane in Figs.~\ref{fig:exclusion_contour_1}--\ref{fig:exclusion_contour_3}. Additionally, the constraints on the standard SI and SD cross-sections (Eq.~\ref{eq:si_sd_cs}) in the $m_\chi-\delta$ plane are presented in Fig.~\ref{fig:xenon_migal_exclusions_exothermic_cs}. 
In particular, from Figs.~\ref{fig:exclusion_contour_1}--\ref{fig:exclusion_contour_3}, it is evident that the Migdal exclusion bounds are overcome by those from nuclear recoil for $\delta \gsim 3$ keV. For this reason in Figs.~\ref{fig:xenon_migal_exclusions_elastic_1}--\ref{fig:xenon_migal_exclusions_exothermic_3} we have only shown exclusion limits for $\delta=0$ and $\delta<0$, omitting $\delta>0$.

In Figs.~\ref{fig:xenon_migal_exclusions_elastic_1}--\ref{fig:xenon_migal_exclusions_exothermic_3}, the Migdal exclusion limits from the XENON1T, SuperCDMS, and DS50 experiments are represented by blue-dashed, cyan-dashed, and green-dashed lines, respectively. The shaded grey region in each plot represents the area excluded by nuclear recoil obtained from the data of a variety of experiments, including XENON1T~\cite{xenon_2018}, PICO60 (using a C$_3$F$_8$~\cite{pico60} and a CF$_3$I target~\cite{pico60_2015}), DS50~\cite{ds50}, LZ~\cite{LZ_2022}, CDEX~\cite{cdex}, CDMSLite~\cite{cdmslite_2017}, COSINE~\cite{cosine}, COUPP~\cite{coupp}, CRESST-II~\cite{cresst_II}, PANDAX-II~\cite{panda_2017}, PICASSO~\cite{PICASSO}, SuperCDMS~\cite{super_cdms_2017}, and XENONnT~\cite{xenonnt}. From these experiments we select the most constraining bound for each combination of $m_\chi$ and $\delta$. The implementation details of the considered experiments are provided in Appendices~\ref{app:Migdal_exp}-\ref{app:nuclear_exp}.

In Figs.~\ref{fig:xenon_migal_exclusions_elastic_1}--\ref{fig:exclusion_contour_3} we separate the interaction operators $\mathcal{O}_{j}$ among SI--type and SD--type. The SI--type operators ($\mathcal{O}_1$, $\mathcal{O}_3$, $\mathcal{O}_{11}$, $\mathcal{O}_{12}$ and $\mathcal{O}_{15}$) are enhanced for heavy targets and are driven either by the $W^{\tau\tau^{\prime}}_{T M}$ or by the $W^{\tau\tau^{\prime}}_{T \Phi^{\prime\prime}}$ nuclear response functions (see Eq.~(\ref{eq:squared_amplitude})). In particular $W^{\tau\tau^{\prime}}_{T M}$ corresponds to the standard SI coupling, proportional to the square of the nuclear mass number; on the other hand, $W^{\tau\tau^{\prime}}_{T \Phi^{\prime\prime}}$ is non vanishing for all nuclei and favors heavier elements with large nuclear shell orbitals not fully occupied. The SD interactions ($\mathcal{O}_4, \mathcal{O}_{6}, \mathcal{O}_{7}, \mathcal{O}_{9}, \mathcal{O}_{10}, \mathcal{O}_{13}, \mathcal{O}_{14}$) vanish when the spin of the nucleus is zero. Among them $\mathcal{O}_4$, $\mathcal{O}_{6}$, $\mathcal{O}_{7}$, $\mathcal{O}_{9}$ and $\mathcal{O}_{10}$, $\mathcal{O}_{14}$ couple to $W^{\tau\tau^{\prime}}_{T \Sigma^\prime}$ and $W^{\tau\tau^{\prime}}_{T \Sigma^{\prime\prime}}$ which are driven by the spin of the target (the sum $W^{\tau\tau^{\prime}}_{T \Sigma^{\prime}}+W^{\tau\tau^{\prime}}_{T \Sigma^{\prime\prime}}$ corresponds to the standard spin-dependent form factor). The $\mathcal{O}_{13}$ operator couples to the $W^{\tau\tau^{\prime}}_{T \tilde{\Phi}^\prime}$, which requires a target spin $j_T > 1/2$. Finally, the two operators $\mathcal{O}_{5}$ and $\mathcal{O}_{8}$ deserve a separate discussion, since, at variance with all other cases, for them velocity-suppressed contributions (proportional to $(v^\perp/c)^2\simeq\mathcal{O}(10^{-6})$) can be of the same order of velocity-independent ones. As a consequence they are driven by a combination of $W^{\tau\tau^{\prime}}_{T \Delta}$ (that is velocity-independent and couples to the nuclear spin) and $W^{\tau\tau^{\prime}}_{T M}$ (velocity suppressed). In particular, in Ar, that has no spin, $W^{\tau\tau^{\prime}}_{T \Delta}$ vanishes and the signal is driven by $W^{\tau\tau^{\prime}}_{T M}$, so it is of the SI--type. On the other hand, in the case of Ge and Xe both $W^{\tau\tau^{\prime}}_{T \Delta}$ and $W^{\tau\tau^{\prime}}_{T M}$ are different from zero: for Ge $W^{\tau\tau^{\prime}}_{T \Delta}$ drives the signal, since the contribution of $W^{\tau\tau^{\prime}}_{T M}$ is negligible due to the velocity suppression, so it can be considered of the SD--type; on the other hand in Xe  the enhancement of $W^{\tau\tau^{\prime}}_{T M}$ due to the large target mass compensates for the velocity suppression, so that it can be considered of the SI--type.
From Figs.~\ref{fig:xenon_migal_exclusions_elastic_1}--\ref{fig:xenon_migal_exclusions_elastic_3} one can see that for $\delta=0$ the Migdal effect allows to probe (exclude) $m_\chi \gsim 30$ MeV, while nuclear recoil reaches $m_\chi \gsim 600$ MeV for SI--type interactions and $m_\chi \gsim 1$ GeV for SD ones. For $\delta=0$  similar conclusions were drawn within the framework of a relativistic effective field theory in~\cite{Tomar:2022ofh}, although in such scenario only a subset of the 14 NR operators $\mathcal{O}_j$ considered here contributes to scattering process at low energy. So the present study complements~\cite{Tomar:2022ofh} by exploring the missing operators. The use of the Migdal effect allows to significantly extend the sensitivity of direct detection searches to lower WIMP masses also thanks to the complementarity of different experiments. 
In particular in Figs.~\ref{fig:xenon_migal_exclusions_elastic_1}--\ref{fig:xenon_migal_exclusions_elastic_3} for SI--type interactions DS50~\cite{ds50_migdal} determines the bounds for $m_\chi \gtrsim 30$ MeV and XENON1T~\cite{XENON_migdal} at higher $m_\chi$. 
On the other hand, since the Ar target has no spin DS50 does not put any bound on SD--type interactions, where, instead, the bounds are driven by SuperCDMS~\cite{supercdms_migdal} at low WIMP mass due to its low 70 eV threshold, while XENON1T~\cite{XENON_migdal} overcomes the SuperCDMS exclusion limit for $m_\chi \geq 100$ MeV.

The exclusion limits for the exothermic scattering are presented for $\delta=-10$ keV in Figs.~\ref{fig:xenon_migal_exclusions_exothermic_1}--\ref{fig:xenon_migal_exclusions_exothermic_3}. 
In this scenario XENON1T drives all the bounds which, compared to the case $\delta=0$, become flat at low WIMP masses and saturate the lower cut on the WIMP mass discussed in Section~\ref{sec:migdal_inelastic} required to comply with the impulse approximation.

Figs.~\ref{fig:exclusion_contour_1}--\ref{fig:exclusion_contour_3} show the bounds on the $c_j$ effective couplings in the $m_\chi-\delta$ plane. The contours show the most constraining bound among XENON1T~\cite{XENON_migdal}, DS50~\cite{ds50_migdal}, and SuperCDMS~\cite{supercdms_migdal}. As in the previous plots the shaded grey regions show the areas excluded by nuclear recoil searches. The dashed green, cyan, and blue lines correspond to the impulse approximation cut $E^{max}_R \geq 50$ meV for Ar, Ge, and Xe respectively, showing that the latter determines the upper edges of the contour plots. In this regime the signal is suppressed and eventually vanishes, so such boundaries represents an approximation of the maximal reach of Migdal searches. Their accurate determination would require to calculate the Migdal expected rate  beyond the impulse approximation. 

\section{Conclusion}
\label{sec:conclusion}
In the present paper we have used the Migdal effect to obtain constraints on the Inelastic Dark Matter scenario at low WIMP mass within the NREFT framework. In order to do so we have included the experimental results from XENON1T, SuperCDMS, and DS50. To calculate the bounds we have assumed one non-vanishing coupling at a time in the effective Hamiltonian of Eq.~(\ref{eq:H}) and an isoscalar interaction, i.e. $c_j^1=0$. 

The result of our analysis is that the use of the Migdal effect allows to significantly extend the sensitivity of direct detection searches to lower WIMP masses, especially for exothermic dark matter, $\delta<0$. 
In particular,  the bounds are driven by XENON1T with the exceptions of the case of a vanishing or very small mass splitting $\delta$. In this case DS50 drives the bound when $m_\chi \leq 0.7$ GeV and the interaction is of the SI--type, while SuperCDMS determines the bound for $m_\chi \leq 0.1$ GeV and the interaction is of the SD--type.
On the other hand, when $\delta\gtrsim 0$ the sensitivity of standard direct detection experiments that measure the recoil energy of the nucleus exceeds that of experiments searching for the Migdal effect. 

\newpage
\section*{Acknowledgments}
The work of S.K. and S.S. is supported by the National Research Foundation of Korea (NRF) funded by the Ministry of Education through the Center for Quantum Space Time (CQUeST) with grant number 2020R1A6A1A03047877 and by the Ministry of Science and ICT with grant number RS-2023-00241757. G.T. is forever grateful for the support and wisdom of his Babuji, whose recent passing has left a lasting impact on his life.

\appendix

\section{Implementations of experimental bounds}
\label{app:experiments}

\subsection{Migdal effect}
\label{app:Migdal_exp}

The main features of the experiments included in our analysis are summarized in Table~\ref{tab:migdal_exps}.

\begin{table}[t!]
\begin{center}
\begin{tabular}{|l|l|l|l|}
\hhline{|-|-|-|-|}
Name & exposure (kg days) & energy bin (keV) & number of events\\ \hline
XENON1T & 22,000 & 0.186 - 3.8 & 49\\
DS50 & 12,500 & 0.083 - 0.106 & 20\\
SuperCDMS & [18.8, 17.5] & 0.07 - 2 & [208, 193]\\
\hline
\end{tabular}
\caption{Main features of the experimental set-ups used in our analysis to constrain the Migdal effect.}
\label{tab:migdal_exps}
\end{center}
\end{table}

\subsubsection{XENON1T}
\label{app:xenon1T_Migdal}
We follow the analysis of Ref.~\cite{XENON_migdal} adopting the single energy bin $0.186~\mbox{keV} \leq E_{det} \leq 3.8~\mbox{keV}$ and an exposure of 22 tonne-days. In particular, in~\cite{XENON_migdal} 61 events were observed, with an expected background of 23.4 events, yielding a $90\%$ C.L. constraint of 49 events.

\subsubsection{DS50}
\label{app:ds50_Migdal}
As observed in~\cite{Tomar:2022ofh}, the Migdal effect analysis in DS50~\cite{ds50_migdal} relies on a profile--likelihood procedure which is difficult to reproduce. However, as pointed out in Section~\ref{sec:analysis}, the shape of the Migdal energy spectrum measured experimentally is fixed by the ionisation probabilities and is approximately the same for all WIMP--nucleon interactions, which only determine the overall normalization. As a result we have used the normalisation of the exclusion plot in~\cite{ds50_migdal}, valid for a standard SI interaction, to obtain the bounds for all the other interaction operators. In particular, to reproduce the exclusion Migdal limit of Fig. 3 in~\cite{ds50_migdal} we adopt the energy bin $0.083~\mbox{keV} \leq E_{EM} \leq 0.106~\mbox{keV}$ with 20 events, and an exposure of $\simeq$12.5 tonne-day.

\subsubsection{SuperCDMS}
\label{app:supercdms_Migdal}
The SuperCDMS collaboration recently conducted a dedicated Migdal analysis~\cite{supercdms_migdal}. They examined two datasets with exposures of 18.8 kg-days and 17.5 kg-days. For both datasets we consider the single energy bin $0.07~\mbox{keV} \leq E_{det} \leq 2~\mbox{keV}$ with 208 and 193 DM events~\cite{supercdms_data}, respectively, and use the energy resolution and efficiency from~\cite{supercdms_data}. In our plots we present the most constraining one between the two data sets.

\subsection{Nuclear recoil}
\label{app:nuclear_exp}

For completeness in Figs.~\ref{fig:xenon_migal_exclusions_elastic_1}--\ref{fig:xenon_migal_exclusions_exothermic_cs} we have included  the regions excluded by experiments searching for nuclear recoil events. We have already provided the details about their implementation in previous analyses. Specifically, for {\bf PANDAX-II} see Appendix B1 of~\cite{sogang_scaling_law_nr}; for {\bf DS50} see Appendix B2 of~\cite{sogang_scaling_law_nr}; for {\bf PICASSO} see Appendix B4 of~\cite{sogang_scaling_law_nr}; for {\bf COUPP} see Appendix B5 of~\cite{sogang_scaling_law_nr}; for {\bf CDEX} see Appendix B3 of~\cite{sogang_scaling_law_nr}; for {\bf CRESST-II} see Appendix B7 of~\cite{sogang_scaling_law_nr}; for {\bf LZ} and {\bf XENON1T} see Appendix A1 of~\cite{halo_independent_sogang}; for {\bf PICO60} see Appendix A2 of~\cite{halo_independent_sogang} for $C_3F_8$ and Appendix A3 of~\cite{halo_independent_sogang} for $CF_3I$; for {\bf SuperCDMS} see Appendix B4 of~\cite{LMI_sogang}; for {\bf CDMSLite} see Appendix B5 of~\cite{LMI_sogang}; for {\bf COSINE} see Appendix B6 of~\cite{LMI_sogang}; for {\bf XENONnT} see Appendix A1 of~\cite{halo_independent_long_range_sogang}.


\bibliographystyle{JHEP} 
\providecommand{\href}[2]{#2}\begingroup\raggedright\endgroup

\end{document}